%% file: Many_2_1_XC_v15_arxiv.tex
\documentclass[draftcls, onecolumn,12pt,twoside]{IEEEtran} 
\usepackage{ifpdf}
\usepackage{cite}
\usepackage[pdftex]{graphicx,color}
\DeclareGraphicsExtensions{.pdf}
\usepackage{bm}
\usepackage[cmex10]{amsmath}
\usepackage{amssymb}
\usepackage{array}
\usepackage{syntonly}

\usepackage{color} 

\usepackage{amsfonts}
\usepackage{enumerate}

\usepackage{url}

\newtheorem{theorem}{Theorem}
\newtheorem{definition}{Definition}
\newtheorem{remark}{Remark}
\newtheorem{lemma}{Lemma}

\newcommand{\bl}{}

\IEEEoverridecommandlockouts 

\begin{document}

\title{On the Gaussian Many-to-One X Channel}

\author{{\large Ranga Prasad$^1$, Srikrishna Bhashyam$^2$, and A.
Chockalingam$^1$}
}

\maketitle

 \begin{abstract}
In this paper, the Gaussian many-to-one X channel, which is a 
special case of general multiuser X channel, is studied. 
In the Gaussian many-to-one X channel, 
communication links exist between all transmitters and one of the receivers,
along with a communication link between each transmitter and its corresponding receiver.  
As per the X channel assumption, transmission 
of messages is allowed on \emph{all} the links of the channel.
This communication model is different from the corresponding 
many-to-one interference channel (IC).
Transmission strategies which 
involve using Gaussian codebooks and treating interference 
from a {\em subset} of transmitters as noise are formulated for the above channel. 
Sum-rate is used as the criterion of optimality for evaluating the strategies. 
Initially, a {\bl $3 \times 3$} many-to-one X channel is considered and   
three transmission strategies are analyzed. 
The first two strategies are shown to achieve sum-rate capacity under
certain channel conditions. For the third strategy, a sum-rate outer bound is derived and 
the gap between the outer bound and the achieved rate is characterized. 
These results are later extended to the $K \times K$ case. 
Next, a region in which the many-to-one X channel can be operated 
as a many-to-one IC without loss of sum-rate is identified. Further, in the above 
region,  it is shown that using Gaussian codebooks and treating 
interference as noise achieves a rate point that is within {\bl $K/2 -1$ bits 
from the sum-rate capacity}. 
Subsequently, some implications of the above results to the Gaussian 
many-to-one IC are discussed. Transmission strategies for the many-to-one IC are formulated 
and channel conditions 
under which the strategies achieve sum-rate capacity are obtained. 
A region where the sum-rate capacity can be 
characterized to within {\bl $K/2-1$ bits} is also identified. Finally, the regions 
where the derived channel conditions are
satisfied for each strategy are illustrated for a $3 \times 3$ many-to-one X channel
and the corresponding many-to-one  IC.
\end{abstract}

{\em {\bfseries keywords:}}
{\em {\footnotesize
Many-to-one interference channel, interference channel, X channel, sum capacity.}}

{\footnotesize 
A part of this work was presented at the IEEE 
International Conference on Communications, Sydney, Australia, June 2014.
}

{\footnotesize $^1$Ranga Prasad and A.~Chockalingam are with the Department of ECE, 
Indian Institute of Science, Bangalore 560012, India 
(e-mail:rprasadn@gmail.com, achockal@ece.iisc.ernet.in).}

{\footnotesize $^2$Srikrishna Bhashyam is with the Department of Electrical 
Engineering, Indian Institute of Technology, Madras, India (e-mail: 
skrishna@ee.iitm.ac.in).} 

\section{Introduction}
\label{sec1}

The interference network is a multi-terminal communication network
introduced by Carleial \cite{Carleial}, consisting of $M$ 
transmitters and $N$ receivers, where each transmitter has an 
independent message for each of the $2^N - 1$ possible non-empty subsets of the receivers.
The multiple access channel (MAC), broadcast channel, interference channel
(IC), and X channel (XC) are all special cases of the interference network.

In the {\bl two-user} interference channel, 
each transmitter communicates an independent message to its corresponding receiver, while the cross channels constitute 
interference at the receivers. The interference channel has been studied 
extensively in literature. Although the capacity region of the IC is unknown, 
several inner and outer bounds for the capacity region and sum-rate 
capacity have been derived in \cite{Kramer, Han, Etkin}. 
In \cite{Veeravalli, Shang, Motahari}, 
sum-rate capacity of the IC is characterized in the low-interference regime: 
a regime where using Gaussian inputs and treating interference as noise is
optimal.

By allowing messages on all the links of the IC, we obtain the X channel, i.e., 
both transmitters have  an independent message for 
each receiver, for a total of four messages in the system.
In this sense, the X channel is a generalization of the IC. 
The best known achievable region is due to 
Koyluoglu, Shahmohammadi, and El Gamal \cite{Koyluoglu}. 
This rate region when specialized to the IC
was shown to reduce to the 
Han--Kobayashi rate region \cite{Han}, which is the 
best known achievable region for the IC. 
The sum-rate capacity result for the Gaussian interference channel in the 
low-interference regime was extended to the Gaussian X channel 
in \cite{Huang}. 

\begin{figure}[t]
\centering
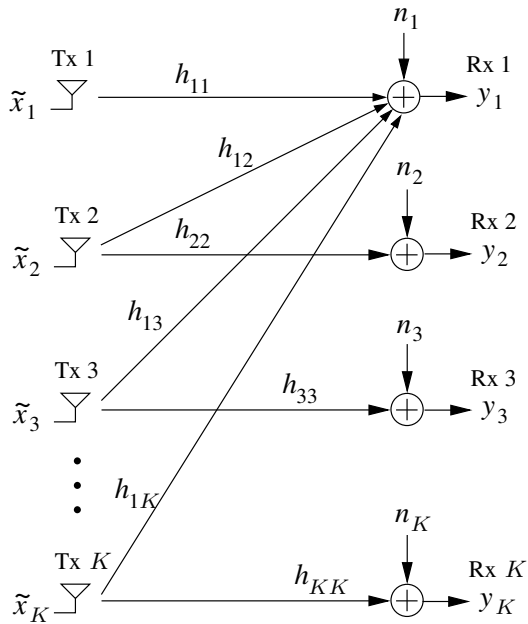
\caption{ $K \times K$ many-to-one X channel system model}
\label{many_2_1_XC_sys_model}
\end{figure}

The many-to-one X channel is a special case of a {\bl $K \times K$ XC, i.e., an XC with $K$ transmitters and $K$ receivers}, 
and can be described as a X channel with ``many-to-one'' connectivity. 
In the many-to-one channel model, communication links exist between 
all transmitters and one of the receivers, say receiver $k$, $k \in \{1, \ldots
K\}$, along with a {\bl direct} communication link
between {\bl transmitter $i$ and receiver $i$, $i = 1, \ldots, K$, $i \ne k$.}
As per the X channel model assumption, transmission of messages is assumed on 
\emph{all} the links of the channel.
The system model for the $K \times K$ many-to-one XC is shown in Fig.
\ref{many_2_1_XC_sys_model},
where we have assumed communication links between all transmitters and 
receiver 1. 
{\bl Thus, for $i = 2, \ldots, K$, each transmitter $i$ has two independent messages, one for receiver $i$, and the other to receiver 1 for a total of $2K-1$ messages in the channel.}
This model has not been studied before. 

The many-to-one interference channel is a special case of the 
many-to-one XC, {\bl where transmitter $i$ is only interested in 
communicating with receiver $i$}, i.e., each
transmitter has only one message. 
The many-to-one 
IC is studied in \cite{Bresler, Veeravalli, Jovicic, Cadambe}. 
In  \cite{Veeravalli, Cadambe}, sum-rate capacity of the many-to-one IC 
is characterized in the low-interference regime. 
In \cite{Bresler}, the capacity region is characterized to 
within a constant number of bits. The generalized degrees of freedom 
of the channel is obtained in \cite{Bresler, Jovicic}.

We study the more general many-to-one X channel with messages 
on all the links. Such a channel could prove useful in the analysis of 
half-duplex relay networks. See \cite{Bama} for examples of such 
networks used in optimization of unicast information flow in multistage 
decode-and-forward relay networks.

\begin{figure}[t]
\centering
\includegraphics[width=0.7\columnwidth]{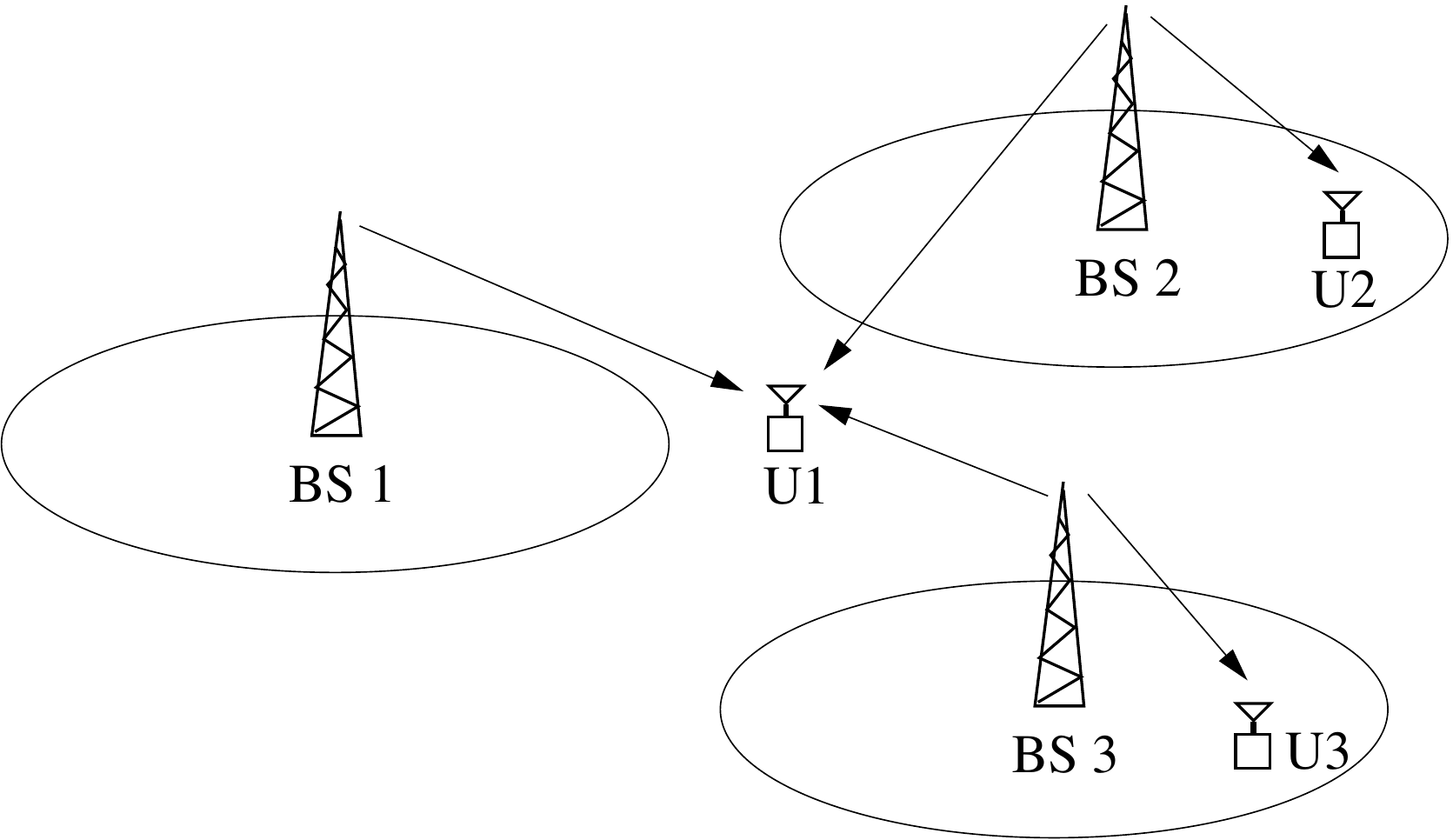}
\caption{Applicability of many-to-one X channel in cellular downlink.}
\label{cell_downlink_illus}
\end{figure}

The many-to-one XC can also occur as a communication model in cellular 
downlink.
Consider the illustration in Fig. \ref{cell_downlink_illus}, 
where user 1 is at the cell edge and receives transmissions from the nearby 
base stations (BS) along with BS 1, while BS 2 and BS 3 simultaneously communicate 
with users 2 and 3, respectively. In order to improve the system throughput, 
all three BSs can communicate independent messages to user 1, provided the 
channel conditions are conducive. The reverse links of this model for 
uplink transmission form the one-to-many X channel studied in \cite{Prasad_WCNC}.

Allowing messages on the cross links leads to some interesting scenarios.
Each transmitter excluding the first, can now make a choice, either transmit to 
its own corresponding receiver, or transmit to receiver 1, or both.
Instead of finding outer and inner bounds to the capacity region 
of the many-to-one XC, we focus on practical transmission scenarios. 
We define the transmission strategies for this channel as follows.

\begin{definition}
In strategy { $\mathcal{M}k$}, transmitter 1 along with $k-1$ other transmitters 
form a MAC at receiver 1, while interference caused by the rest 
of the  transmitters is treated as noise, $k = 1, 2, \ldots, K$. 
All transmitters use Gaussian codebooks.
\end{definition}

In Table \ref{table_strat_many_2_1}, we list all possible strategies as per the 
above definition for $K=3$. Thus, in strategy { $\mathcal{M}1$}, interference 
caused by transmitters 2 and 3 at receiver 1 is treated as noise, while in 
strategy { $\mathcal{M}3$}, 
receiver 1 does not experience any interference.

\begin{table}[t]
\centering
{
\renewcommand{\arraystretch}{2}
\renewcommand{\tabcolsep}{0.34cm}
\begin{tabular}{|c|p{0.79\columnwidth}|}
\hline
\normalsize No. & \normalsize \hspace{3cm}Strategy \\
\hline
 \normalsize $\mathcal{M}1$ & \normalsize All transmitters transmit to their
corresponding receivers and interference
at receiver 1 is treated as noise. \\
\hline
 \normalsize $\mathcal{M}2$ & \normalsize Transmitter 1 and either transmitter 2 
 or transmitter 3 form a MAC at receiver 1, while the interference from the 
 other transmitter is treated as noise. \\
\hline
 \normalsize $\mathcal{M}3$ & \normalsize All transmitters form a MAC at receiver 1.
\\
\hline
\end{tabular}
}
\caption{Transmission strategies for a $ 3 \times 3$ many-to-one XC}
\label{table_strat_many_2_1}
\end{table}

\begin{figure}[t]
\centering
\includegraphics[width=0.78\columnwidth]{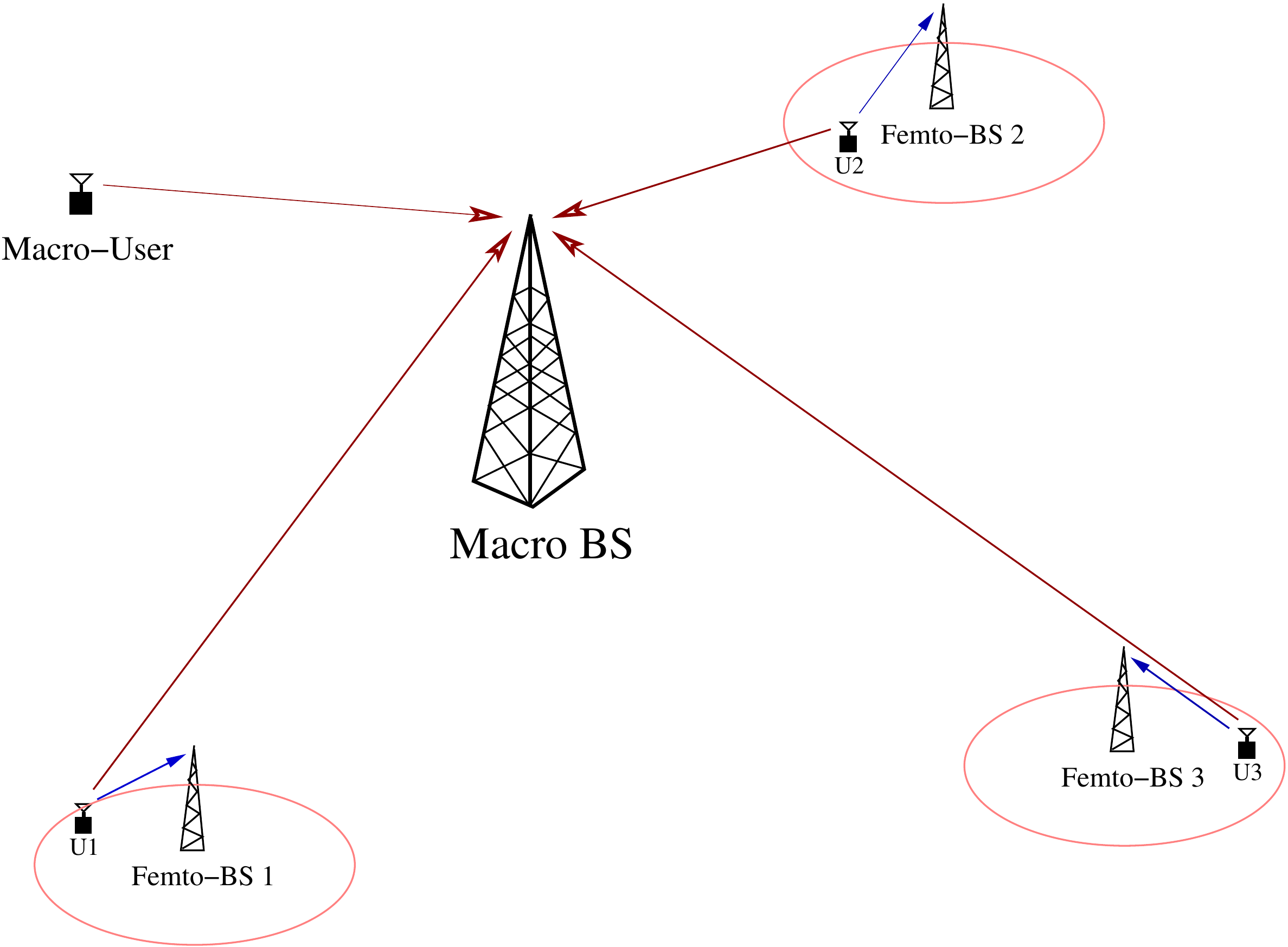}
\caption{Modeling of uplink transmissions in a heterogeneous network (HetNet) 
with macro-BS and femto-BSs as a many-to-one X channel.}
\label{small_cell_uplink_illus}
\end{figure}

The analysis of specific transmission strategies is also motivated by applications 
to small cell networks. 
Small cells encompassing femtocells, picocells, and microcells, are used by 
mobile service providers to increase network capacity and/or extend the service 
coverage area. Consider the illustration in Fig. \ref{small_cell_uplink_illus}, where 
some femto-BSs along with their corresponding users within a small coverage area 
co-exist in a macro cell consisting of macro users served by the macro BS. 
To increase the service reliability and throughput, the users can
either communicate with the femto-BS or with the macro-BS. This communication model also 
results in the many-to-one X channel.

Small cells are seen as an effective means to achieve 3G data off-loading, and 
many mobile service providers consider small cells as a vital element for managing 
LTE Advanced spectrum more efficiently compared to using just macrocells.  
It is in this context that the knowledge of the optimality of different transmission 
strategies that the users can employ becomes valuable. 
Femto, pico and micro cells are also used to motivate a slightly similar channel model 
studied in \cite{Zhu}, where a MAC generates interference for a single user 
uplink transmission.
We note that the many-to-one IC was also motivated by considering a similar 
scenario where 
multiple short-range peer-to-peer communications create interference for a 
long-range receiver \cite{Bresler, Jovicic}.

We use a $3 \times 3$ many-to-one XC to evaluate the different strategies.
The sum-rate at all the receivers is used as the criterion for 
optimality. 
In general, we use genie-aided bounding techniques to derive the sum-rate 
capacity results in this paper. Specifically, for certain 
strategies we make use of the concepts of 
\emph{useful  genie} and \emph{smart  genie} introduced in \cite{Veeravalli}.
A genie is said to be useful if it results in a genie-aided channel 
whose sum-rate capacity is achieved by Gaussian inputs, while 
a smart genie is one which does not increase the sum-rate when Gaussian inputs 
are used \cite{Veeravalli}. In \cite{Veeravalli}, the genie-aided bounding technique 
is used to identify the regime under which all the interference can be treated as noise. 
In our work, we use this technique for scenarios where interference from a {\em subset} 
of transmitters is treated as noise.
We show that strategies $\mathcal{M}1$ and $\mathcal{M}2$
achieve sum-rate capacity under certain channel conditions. 
For strategy { $\mathcal{M}3$}, we characterize the gap between the 
achievable sum-rate of the strategy and a sum-rate outer bound. 
Later, we extend these results to the $K \times K$ case.

Next, we identify a region in which the many-to-one XC can be operated as a 
many-to-one IC without loss of sum-rate and show that using Gaussian codebooks and treating interference as noise achieves a rate point that is within $K/2-1$ bits from the 
sum-rate capacity. In the last part of the paper, we observe some implications 
of the above results for the many-to-one IC. Firstly, we note that strategies 
similar to the ones defined above can be considered for the many-to-one IC as well. 
These involve a combination of partial interference cancellation and treating the 
rest of the interference as noise. We derive the sum-rate optimality of these 
strategies under certain channel conditions. Secondly, we identify a region for 
the many-to-one IC where the sum-rate capacity can be characterized to $K/2-1$ bits.

{\bl 
In this paper, we restrict ourselves to the many-to-one topology. In general, for the fully
connected $K \times K$ XC, obtaining regions where conventional transmission strategies are sum-rate capacity optimal is difficult. However, some gap-to-capacity results have recently been obtained in \cite{Geng_IC, Geng_XC_ISIT, Geng_XC, Niesen}. In  \cite{Geng_IC}, channel conditions under which treating interference as noise at the receivers (strategy $\mathcal{M}1$) achieves the entire channel
capacity region of the $K$-user Gaussian interference channel to within a constant gap are obtained. This result is extended to the $K \times K$ XC in \cite{Geng_XC_ISIT, Geng_XC} to show that under the same channel conditions, treating interference as noise is optimal in terms of sum-rate capacity up to a constant gap. In \cite{Niesen}, a constant gap capacity approximation for the $2 \times 2$ XC subject to an outage set has been obtained.
}

The rest of this paper is organized as follows. 
The system model is presented in Section \ref{sec_sys_mod}.
In Section \ref{sec_opt_strategies}, we consider the $3 \times 3$ many-to-one XC  
and analyze the different strategies defined earlier.  
These results are extended to the $K \times K$ case in Section
\ref{sec_exten_K_tx}. Some implications of the above results for the Gaussian 
many-to-one IC are discussed in Section \ref{sec_gauss_m21_ic}.
Numerical results and illustrations regarding the 
optimality of the strategies are presented in Section \ref{sec_num_res}. 
Conclusions are presented in Section \ref{sec_concl}.

\section{System Model}
\label{sec_sys_mod}

As shown in Fig. \ref{many_2_1_XC_sys_model}, the 
many-to-one XC {\bl with $K$ transmitters and $K$ receivers} is described by the 
following input-output equations
\begin{eqnarray}
y_1 &=& h_{11} \, \tilde{x}_1 + \sum_{j=2}^K h_{1j} \, \tilde{x}_j + \, n_1
\label{out_eqn_1} \\ 
y_i &=& h_{ii} \, \tilde{x}_i  + \, n_i, \qquad \qquad\qquad i  = 2, 3, \ldots, K,
\label{out_eqn_2}
\end{eqnarray}
where  $\tilde{x}_t$ is\footnote{We use
the following notation: lowercase letters for scalars, boldface lowercase
letters
for vectors, and calligraphic letters for sets.
$[\cdot]^T$ denotes the transpose
operation,
$\mbox{trace}(\cdot)$ denotes the trace operation, and ${\mathbb E}\{\cdot\}$ 
denotes the expectation operation. {\bl $\big\|{\bf x} \big\|_2$ denotes the $l_2$ norm of 
the row or column vector ${\bf x}$.} }
 the transmitted symbol by transmitter $t$, $h_{rt}$ denotes 
the channel coefficient from transmitter $t$ to receiver $r$,
and $n_r$ is the additive Gaussian noise at receiver $r$.
$h_{ii}$, $i = 2, \ldots, K$, are the direct channels, while $h_{1i}$ 
are the cross channels.
The additive noise $n_r$ is a zero mean Gaussian 
random variable with unit variance, i.e., $n_{r} \sim \mathcal{N}(0, 1)$, $r =
1, 2, \ldots, K$.

\begin{figure}[t]
\centering
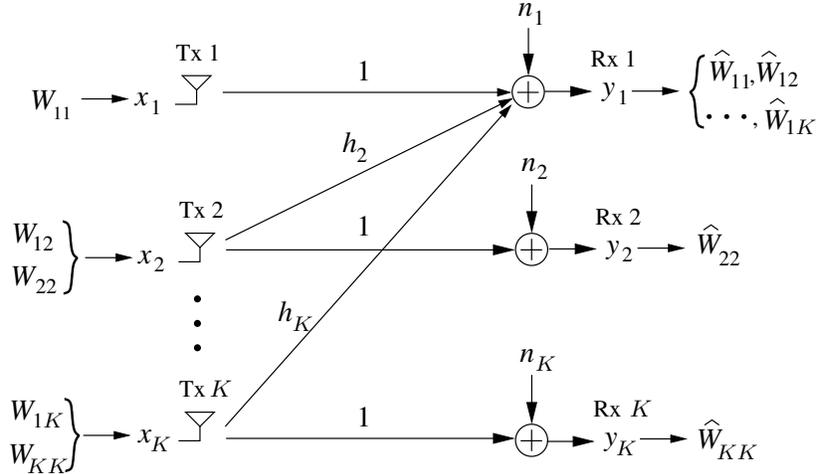
\caption{Many-to-one X channel with $K$ transmitters in standard
form.}
\label{many_2_1_XC_Ktx_std_form}
\end{figure}

{\bl
\subsection{$K \times K$ Many-to-one X channel in standard form}

The $K \times K$ many-to-one XC can be 
written in standard form (see Fig. \ref{many_2_1_XC_Ktx_std_form}), i.e.,
\begin{eqnarray}
y_1 &=& x_1 + \sum_{j=2}^K h_{j} \, {x}_j + \, n_1
\label{m21_xc_std_out_eqn_1_Ktx} \\ 
y_i &=& {x}_i  + \, n_i, \qquad i  = 2, 3, \ldots, K,
\label{m21_xc_std_out_eqn_2_Ktx} 
\end{eqnarray}
where we have used $h_j = h_{ij}\,/\, h_{jj}$, 
$x_i  =  h_{ii} \, \tilde{x}_i$, and 
$P_i = |h_{ii}|^2 \tilde{P}_i$ are the new power constraints \cite{Carleial}.

As shown in Fig. \ref{many_2_1_XC_Ktx_std_form}, the $K \times K$ many-to-one
XC has $2 K -1$ independent
messages, i.e., \{$W_{11}$, $W_{12}$, $W_{22}$, $W_{13}$, $W_{33}$, \ldots,
$W_{1K}$, $W_{KK}$\}, where $W_{ij}$ is
the message transmitted from transmitter $j$ to receiver $i$.

We assume that the transmitter communicates the intended messages in $n$ channel uses. 
For a given block length $n$, we define a $\big(n, R_{11}\big)$ codebook at transmitter 1, and $\big(n, R_{ii}, R_{1i}\big)$ codebook at transmitter $i$, $i = 2, \ldots, K$, as follows:
\begin{enumerate}
\item Transmitter 1 communicates message $W_{11} \in \mathcal{W}_{11} = \{1, \ldots, 2^{n R_{11}}\}$, while Transmitter $i$ communicates messages $W_{ii} \in \mathcal{W}_{ii} = \{1, \ldots, 2^{n R_{ii}}\}$ and $W_{1i} \in \mathcal{W}_{1i} = \{1, \ldots, 2^{n R_{1i}}\}$, $i = 2, \ldots, K$.

\item An encoding function $f_1(\cdot)$ at transmitter $1$ maps the message $W_{11}$ to the transmitted codeword ${\bf x}_1^n = (x_{11}, x_{12}, \ldots, x_{1n})$,  $f_1 \,:\, (W_{ii}, W_{1i}) \rightarrow {\bf x}_1^n$ for each $W_{11} \in \mathcal{W}_{11}$. Similarly, for transmitter $i$, an encoding function $f_i(\cdot)$ maps the messages to the transmitted codewords, $f_i \,:\, (W_{ii}, W_{1i}) \rightarrow {\bf x}_i^n$ for each $(W_{ii}, W_{1i}) \in \mathcal{W}_{ii} \times \mathcal{W}_{1i}$, for $i = 2, \ldots, K$. 

\item The codewords in each codebook must satisfy the average power constraint $\frac{1}{n}\big\| {\bf x}_i^n \big\|_{2}^2 \le P_i$ at transmitter $i = 1, \ldots, K$.

\item Receiver $i$ observes the channel outputs ${\bf y}_i^n =  (y_{i1}, y_{i2}, \ldots, y_{in})$ and uses a decoding function $\phi_k(\cdot)$ at receiver $k$ which maps the received symbols to an estimate of the message: $\phi_1({\bf y}_1) = (\hat{W}_{11}, \hat{W}_{12}, \ldots, \hat{W}_{1K})$ and $\phi_k({\bf y}_k) = \hat{W}_{kk}$ for $k = 2, \ldots, K$.
\item The average probability of error at receiver $k$, $P_{e, k}^{(n)}$  is given by 
\begin{eqnarray}
P_{e,1}^{(n)} & = &  \mathbb{E} \left[\text{Pr} \left(\big(\hat{W}_{11}, \hat{W}_{12}, \ldots, \hat{W}_{1K} \big) \ne ({W}_{11}, {W}_{12}, \ldots, {W}_{1K}\big) \right) \right] \nonumber \\
P_{e, k}^{(n)} & = & \mathbb{E} \left[\text{Pr}\big(\hat{W}_{kk} \ne {W}_{kk} \big)\right], \qquad k = 2, \ldots, K, \nonumber
\end{eqnarray}
where the expectation is taken with respect to the random choice of the transmitted messages.
\end{enumerate}

We say that the rate vector $(R_{11}, R_{12}, \ldots, R_{1K}, R_{22}, \ldots R_{KK})$ is achievable for the $K \times K$ many-to-one XC if there exists a
 $\big(n, R_{11}\big)$ codebook at transmitter 1 satisfying the power constraint $P_1$, and $\big(n, R_{ii}, R_{1i}\big)$ codebook at transmitter $i$ satisfying the power constraint $P_i$, $i = 2, \ldots, K$, and decoding functions $(\phi_1(\cdot), \ldots, \phi_K(\cdot))$, 
such that the average decoding error probabilities $(P_{e, 1}^{(n)}, \ldots, P_{e, K}^{(n)})$ go to zero as block length $n$ goes to infinity. The capacity region is defined as the closure of
the set of all achievable rate vectors $(R_{11}, R_{12}, \ldots, R_{1K}, R_{22}, \ldots R_{KK})$ and is denoted by $\mathcal{C}$. Then the sum-rate capacity $S$ of the $K \times K$ many-to-one XC is defined as
\begin{eqnarray}
S &  = & \max_{(R_{11}, R_{12}, \ldots, R_{1K}, R_{22}, \ldots R_{KK}) \, \in \, \mathcal{C}} \Big( R_{11} + \sum_{i = 2}^K (R_{ii} + R_{1i}) \Big).
\nonumber 
\end{eqnarray}

By Fano's inequality, we have
\begin{eqnarray}
H(W_{ii} \,|\, {\bf y}_i^n ) & \le & n \epsilon_n, \quad i = 1, \ldots, K, \nonumber \\
H(W_{1j} \,|\, {\bf y}_1^n ) & \le & n \epsilon_n, \quad j = 2, \ldots, K,
\label{many_2_1_fano_ineq_Ktx}
\end{eqnarray}
where $\epsilon_n \rightarrow 0$ as $n \rightarrow \infty$.

Next, in Lemma \ref{lemma_dec_msg_rx_i} below, we show that the $K \times K$ many-to-one XC is degraded under specific channel conditions. This lemma will later be used to prove the decodability of message sets at the receivers. In order for the result to be applicable to a more general case, we assume that the noise variance at each receiver is $\sigma_i^2$, $i = 1, \ldots, K$.

\begin{lemma}
For the $K \times K$ many-to-one XC in standard form shown in Fig. \ref{many_2_1_XC_Ktx_std_form} with noise variance $\sigma_i^2$ at receiver $i$, 
if $h_i^2 \, \sigma_i^2  \le \sigma_1^2$, $i = 2, \ldots, K$, then $y_1$ is a degraded version of $y_i$ with respect to message $W_{1i}$ and hence $H(W_{1i} \, | \, {\bf y}_i^n) \le n \epsilon_n$, where $\epsilon_n \rightarrow 0$ as $n \rightarrow \infty$. 
This implies that message $W_{1i}$ is decodable at receiver $i$. 
Furthermore, $H(W_{1i}, W_{ii} \, | \, {\bf y}_i^n) \le 2 n \epsilon_n$. 
\label{lemma_dec_msg_rx_i}
\end{lemma}

\begin{IEEEproof}
At receiver 1, we have $y_1 = x_1 + \sum_{j=2}^K h_{j} \, {x}_j + \, n_1$, and at receiver $i$, we have $y_i = x_i + n_i$. Define $\tilde{y}_1 = h_i x_i + n_1$ and $y_1' = \tilde{y}_1 / h_i =  x_ i + n_1'$, where $n_1' = n_1  / h_i$. If $\sigma_i^2 \le \sigma_1^2 / h_i^2$, 
we note that the noise variance of $n_1'$ is higher than that of $n_i$. Hence $y_1'$ is a stochastically degraded version of the signal $y_i$ received at receiver $i$. Thus, from the data processing inequality, we have $I(W_{1i} \, ; \, {\bf y}_i^n) \ge I(W_{1i} \, ;  \, {\bf y}_1'^n)$. Since scaling the output of a channel does not affect its capacity, we have 
$I(W_{1i} \, ; \, {\bf y}_i^n) \ge I(W_{1i} \, ;  \, \tilde{\bf y}_1^n)$. 
Therefore, 
\begin{eqnarray}
H(W_{1i} \,|\, {\bf y}_i^n) & \le & H(W_{1i} \,|\, \tilde{\bf y}_1^n)
\nonumber\\
& \stackrel{(a)}{=} & H(W_{1i} \,|\, \tilde{\bf y}_1^n,  {\bf x}_1^n, 
\ldots, {\bf x}_{i-1}^n, {\bf x}_{i+1}^n, \ldots, {\bf x}_K^n). \nonumber \\
& = & H(W_{1i} \,|\, {\bf y}_1^n,  {\bf x}_1^n, 
\ldots, {\bf x}_{i-1}^n, {\bf x}_{i+1}^n, \ldots, {\bf x}_K^n). \nonumber \\
& \stackrel{(b)}{\le} & H(W_{1i} \,|\, {\bf y}_1^n) \nonumber \\ 
& \stackrel{(c)}{\le} & n \epsilon_n, \label{h_w1i_y_i}
\end{eqnarray}
where $(a)$ follows since $\big({\bf x}_1^n, 
\ldots, {\bf x}_{i-1}^n, {\bf x}_{i+1}^n, \ldots, {\bf x}_K^n\big)$ are independent of $W_{1i}$ and
$\tilde{\bf y}_1^n$, $(b)$ follows from the fact that removing conditioning does not reduce the conditional entropy, and $(c)$ follows from \eqref{many_2_1_fano_ineq_Ktx}.
Thus, we conclude that $W_{1i}$ is decodable at 
receiver $i$ when $h_i^2 \, \sigma_i^2  \le \sigma_1^2$. Note that in this case
\begin{eqnarray}
H(W_{1i}, W_{ii} \, | \, {\bf y}_i^n) 
& = & H(W_{1i} \, | \, {\bf y}_i^n) + H(W_{ii} \, | \, {\bf y}_i^n, W_{1i}) \nonumber \\
& {\le} &  H(W_{1i} \, | \, {\bf y}_i^n) + H(W_{ii} \, | \, {\bf y}_i^n)   \nonumber \\
& \le & 2 n \epsilon_n, \label{m21_dec_x_3_at_yi}
\end{eqnarray}
where \eqref{m21_dec_x_3_at_yi} follows from \eqref{many_2_1_fano_ineq_Ktx} 
and \eqref{h_w1i_y_i}. As $n \rightarrow \infty$, $\epsilon_n \rightarrow 0$.  This shows that $(W_{ii}, W_{1i})$ are decodable at receiver $i$.
\end{IEEEproof}
}

\subsection{$3 \times 3$ Many-to-one X channel}

In order to analyze the strategies, we first consider the $3 \times 3$ many-to-one XC 
since the $2 \times 2$ case results in the Z channel.
The Z channel is obtained from the many-to-one XC by retaining 
only the first two transmitters and removing the rest. 
In this way, the many-to-one XC can be considered as one 
possible generalization of the Z channel. 
The Z channel has been studied in \cite{Liu, Chong}.

\begin{figure}[t]
\centering
\includegraphics[width=0.65\columnwidth]{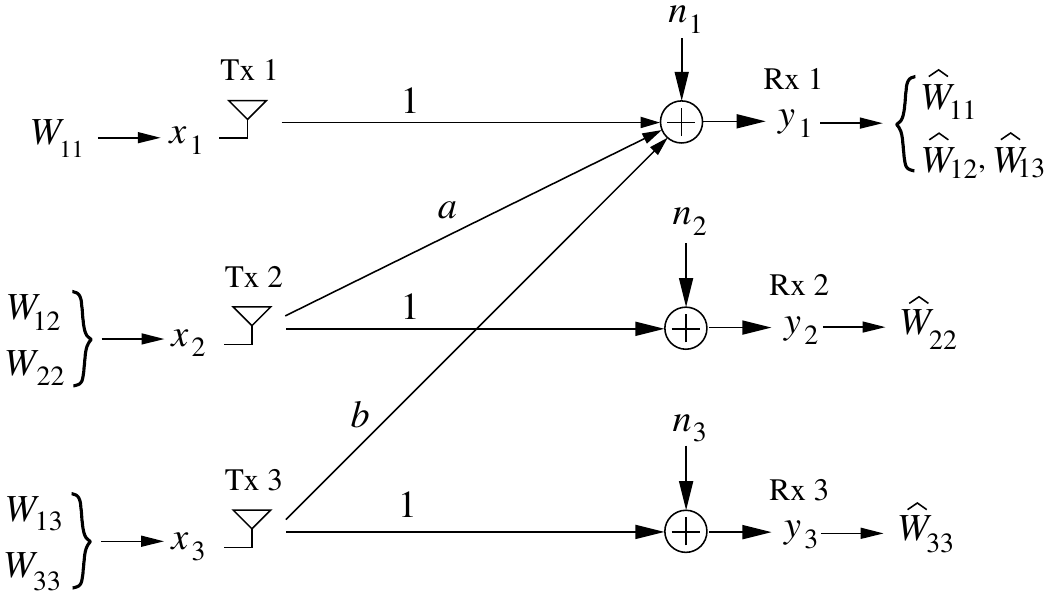}
\caption{Many-to-one X channel with $3$ transmitters in standard
form.}
\label{many_2_1_3tx_std_form}
\end{figure}

The $3 \times 3$ many-to-one XC channel can be 
written in standard form (See Fig. \ref{many_2_1_3tx_std_form}), i.e.,
\begin{eqnarray}
y_1 &=& x_1 + a x_2 + b x_3 + \, n_1
\label{m21_xc_std_out_eqn_1_3tx} \\ 
y_2 &=& \, x_2  + \, n_2 \label{std_out_eqn_2} \\
y_3 &=& \, x_3  + \, n_3,
\label{m21_xc_std_out_eqn_3_3tx}
\end{eqnarray}
{\bl where we have used $h_2 = a$ and $h_{3} = b$.}

As shown in Fig. \ref{many_2_1_3tx_std_form}, the $3 \times 3$ many-to-one XC has five independent 
messages, $W_{11}$, $W_{12}$, $W_{13}$, $W_{22}$ and $W_{33}$,  where $W_{ij}$ is 
the message transmitted from transmitter $j$ to receiver $i$.

Our motivation for considering the $3 \times 3$ many-to-one XC first, instead of 
directly analyzing $K \times K$ case stems from three perspectives: 
(i) ease of presentation, (ii) understanding the proof techniques without 
cumbersome notational details, (iii) better visualization of the regions where the 
strategies are optimal (as seen in the numerical results 
presented in Section \ref{sec_num_res}). 

\section{Analysis of Different Strategies for the $3 \times 3$ Many-to-One XC}

\label{sec_opt_strategies}

We introduce some terminology useful in deriving the results  in this section.
Let ${\bf y}_i^n$ denote the vector of received symbols of length $n$ at 
receiver $i$. Let ${\bf x}_i^n$ denote the $n$ length 
vector of transmitted symbols at transmitter $i$. 
By Fano's inequality, we have
\begin{eqnarray}
H(W_{ii} \,|\, {\bf y}_i^n ) & \le & n \epsilon_n, \quad i = 1, 2, 3 \nonumber \\
H(W_{1j} \,|\, {\bf y}_1^n ) & \le & n \epsilon_n, \quad j = 2, 3, 
\label{many_2_1_fano_ineq}
\end{eqnarray}
where $\epsilon_n \rightarrow 0$ as $n \rightarrow \infty$.

Before we proceed to analyze the various strategies, we provide a 
restatement of Lemma 5 in \cite{Veeravalli}, in a form that is easier to apply
to the many-to-one X channel. 
We make use of the following lemma to bound the 
sum-rate of the many-to-one XC in some cases.

\begin{lemma}
Let ${\bf w}_i^n$ be a sequence with 
average power constraint $\text{trace}(\mathbb{E}({\bf w}_i^n {\bf w}_i^{nT}))
\le n P_i$.
Let ${\bf n}_i^n$, {\bl $i \ne 1$},  be a random vector with components that are
distributed as independent $\mathcal{N}(0,1)$ random variables. {\bl Let ${\bf n}_1^n$ denote a random vector with components that are distributed as  independent $\mathcal{N}(0, \sigma^2)$ random variables.} Assume that ${\bf w}_i^n$ are independent 
of each other and also independent of ${\bf n}_i^n$. Let $w_{iG} \sim
\mathcal{N}(0, P_i)$. For some
constants $c_i$, we have 
\begin{eqnarray}
\sum_{i=1}^{K} h({\bf w}_i^n + {\bf n}_i^n) - h\Big(\sum_{i=1}^{K} c_i \,{\bf w}_i^n +
{\bf n}_1^n\Big) & \le &  n \sum_{i=1}^{K} h({w}_{iG} + {n}_i) - n h\Big( \sum_{i=1}^{K}
c_i \,{w}_{iG} + {n}_{1}\Big),\quad 
\label{lemma_li_eqn}
\end{eqnarray}
when $\sum_{i=1}^{K} c_i^2 \le {\bl \sigma^2}$ and equality is achieved if ${\bf w}_i^n =
{\bf w}_{iG}^n$, where
${\bf w}_{iG}^n$ denotes a complex random vector with components that are i.i.d
$\mathcal{N}(0, P_i)$.
\label{lemma_li}
\end{lemma}

\begin{IEEEproof}
Let ${\bf t}^n_i =  c_i ({\bf w}_i^n + {\bf n}_i^n)$. The  
left-hand side of \eqref{lemma_li_eqn} can now be written as 
\begin{eqnarray*}
\sum_{i=1}^{K} h({\bf t}^n_i) - h\Big(\sum_{i=1}^{K} {\bf t}^n_i + \tilde{\bf
n}_1^n\Big) + n \sum_{i=1}^{K} \log c_i,
\end{eqnarray*}
where $\tilde{\bf n}_1^n$ is a random vector with components that are
i.i.d
$\mathcal{N}\big(0, {\bl \sigma^2} - \sum_{i=1}^{K} c_i^2\big)$. The final result follows by
applying
Lemma 5 in \cite{Veeravalli}, i.e.,
{\setlength{\arraycolsep}{2pt}
\begin{eqnarray*}
\sum_{i=1}^{K} h({\bf t}^n_i) - h\Big(\sum_{i=1}^{K} {\bf t}^n_i + \tilde{\bf
n}_1^n\Big) & \le &  \; n \sum_{i=1}^{K} h({t}_{iG}) - n h\Big(\sum_{i=1}^{K}
{t}_{iG} + \tilde{n}_1\Big),
\end{eqnarray*}
}where $t_{iG} = c_i ({w}_{iG} + {n}_i)$ and equality is achieved if ${\bf
w}_i^n = {\bf w}_{iG}^n$.
Since the variance of $\tilde{n}_1$ cannot be negative, we have the condition
$\sum_{i=1}^{K} c_i^2 \le {\bl \sigma^2}$. \end{IEEEproof}

\subsection{Optimality of Strategy $\mathcal{M}1$}

\begin{table}[t]
\centering
{
\renewcommand{\arraystretch}{1.2}
\renewcommand{\tabcolsep}{.2cm}
\begin{tabular}{|>{\normalsize}c|>{\normalsize}c|>{\normalsize}c|}
\hline
 Transmitter/  &  Transmitted & Decoded  \\
Receiver index & messages & messages \\
\hline
1 & $W_{11}$ & $\widehat{W}_{11}$ \\
\hline
2 & $W_{22}$ & $\widehat{W}_{22}$ \\
\hline
3 & $W_{33}$ & $\widehat{W}_{33}$ \\
\hline
\end{tabular}
}
\caption{Transmitted and Decoded messages for strategy $\mathcal{M}1$}
\label{table_strategy_M1}
\end{table}

{\bl The transmitted and decoded messages in strategy $\mathcal{M}1$ are illustrated in Table \ref{table_strategy_M1}. In strategy $\mathcal{M}1$, we are interested in a region where sum-rate capacity is achieved by using Gaussian codebooks and treating interference as noise. This is usually referred to as the \emph{low-interference} or the 
\emph{noisy-interference} regime in the interference channel 
literature. In strategy $\mathcal{M}1$, cross messages in the channel are not utilized, i.e., $W_{12} = W_{13} =  \phi$.
We characterize the noisy-interference sum-rate capacity in the  
following theorem.}

\begin{theorem}
{\bl
For the $3 \times 3$ Gaussian many-to-one XC, strategy $\mathcal{M}1$ achieves sum-rate capacity if }
\begin{eqnarray}
a^2 + b^2 & \le &  1, \label{m21_3tx_low_interf_cond}
\end{eqnarray}
and the sum-rate capacity is given by 
\begin{eqnarray}
 S & = & 0.5 \log \bigg(1 + \frac{P_1}{1 + a^2 P_2 +  b^2 P_3} \bigg ) + 0.5\log(1 + P_2) + 0.5\log(1 + P_3).
\end{eqnarray}
\label{thm_low_interf}
\end{theorem}

\begin{IEEEproof}
{\bl If $b^2 \le 1$, from Lemma \ref{lemma_dec_msg_rx_i}, we have $H(W_{13} \,|\, {\bf y}_3^n) \le n \epsilon_n$ and $H(W_{13}, W_{33} \, | \, {\bf y}_3^n) \le 2n \epsilon_n$. Similarly, if $a^2 \le 1$, $W_{12}$ is decodable at receiver 2, i.e., $H(W_{12} \,|\, {\bf y}_2^n) \le n \epsilon_n$ and $H(W_{12}, W_{22} \, | \, {\bf y}_2^n) \le 2n \epsilon_n$.} 

Now, assume $a^2 \le 1$ and $b^2 \le 1$.
The sum-rate can be bounded as follows:
\begin{eqnarray}
n S & \le & H(W_{11}) + H(W_{12}, W_{22}) + H(W_{13}, W_{33}) \nonumber \\
& = & {\bl I(W_{11} \, ; \, {\bf y}_1^n) + \sum_{i=2}^3 I(W_{1i}, W_{ii} \, ; \, {\bf y}_i^n) + H(W_{11} \, | \, {\bf y}_1^n) + \sum_{i=2}^3 H(W_{1i}, W_{ii} \, | \, {\bf y}_i^n) }
\nonumber \\
& \le & {\bl I({\bf x}_1^n \, ; \, {\bf y}_1^n) + \sum_{i=2}^3 I({\bf x}_i^n \, ; \, {\bf y}_i^n) + H(W_{11} \, | \, {\bf y}_1^n) + \sum_{i=2}^3 H(W_{1i}, W_{ii} \, | \, {\bf y}_i^n)}
\nonumber \\
& \stackrel{(a)}{\le} & h({\bf y}_1^n)  - h(a\;\! {\bf x}_2^n + b\;\! {\bf x}_3^n + {\bf n}_1^n) + h({\bf x}_2^n + {\bf n}_2^n) -  h({\bf n}_2^n) 
+   h({\bf x}_3^n + {\bf n}_3^n)   -  h({\bf n}_3^n) + 5 \epsilon_n \nonumber\\
& \stackrel{(b)}{\le}  & n h({y}_{1G}) +  n h({x}_{2G} + {n}_2 ) + n h({x}_{3G} + {n}_3 )    
 - n h(a\;\! {x}_{2G} + b\;\! {x}_{3G} + { n}_{1}) - nh(n_2) 
\nonumber\\ 
&&-  nh(n_3) + 5 \epsilon_n \label{app_lem1_low_int} \nonumber \\
& = & n I({x}_{1G} ; {y}_{1G}) + n I({x}_{2G} ; {y}_{2G}) + n I({x}_{3G} ; {y}_{3G}) + 5 \epsilon_n,  \nonumber
\end{eqnarray}
where $x_{iG} \sim \mathcal{N}(0, P_i)$, $y_{iG}$ denotes $y_i$ 
with $x_j = x_{jG}$, $\forall \,i, j$, $(a)$ follows from \eqref{many_2_1_fano_ineq}, and from the application of Lemma \ref{lemma_dec_msg_rx_i}, 
and in $(b)$, we have used Lemma \ref{lemma_li} to bound the term
$h({\bf x}_2^n + {\bf n}_2^n) +  h({\bf x}_3^n + {\bf n}_3^n)  
 - h(a {\bf x}_2^n + b {\bf x}_3^n + {\bf n}_1^n)$, under the condition $a^2 + b^2 \le 1$.
As $n \rightarrow \infty$, $\epsilon_n \rightarrow 0$, and we have 
\begin{eqnarray}
S & {\le}  & 0.5\log \bigg(1 + \frac{P_1}{1 + a^2 P_2 +  b^2 P_3} \bigg ) + \sum_{i=2}^3 0.5 \log(1 + P_i).  \quad
\label{sum_bnd_strat_M1}
\end{eqnarray} 
This sum-rate bound can be achieved using strategy $\mathcal{M}1$. {\bl We observe that the  sum-rate bound in \eqref{sum_bnd_strat_M1} is also achievable in the $3 \times 3$ many-to-one IC by using Gaussian inputs and treating interference at receiver 1 as noise. Note that in the $3 \times 3$ many-to-one IC, the cross messages $W_{12}$ and $W_{13}$ are absent. Since the many-to-one IC is a special case of the many-to-one XC, this shows 
that the presence of cross messages does not improve the 
sum-rate when $a^2 + b^2 \le 1$. This means that we can set $W_{12} = W_{13} = \phi$ in the $3 \times 3$ many-to-one XC without loss of sum-rate.}
\end{IEEEproof}
\begin{remark}
Theorem \ref{thm_low_interf} was proved for the many-to-one 
interference channel in \cite[Theorem 4]{Veeravalli} using 
genie aided bounding techniques. 
The low-interference regime for the discrete memoryless 
many-to-one interference channels is proved in \cite{Cadambe}.
We also note that the result in \cite{Veeravalli} 
is a special case of a more general 
result in \cite[Theorem 3]{Shang_1}, 
where the sum-rate capacity of a $K$-user Gaussian interference channel  is characterized 
in the noisy-interference regime.
\end{remark}

\subsection{Optimality of Strategy $\small \mathcal{M}2$}

\begin{table}[t]
\centering
{
\renewcommand{\arraystretch}{1.2}
\renewcommand{\tabcolsep}{.2cm}
\begin{tabular}{|>{\normalsize}c|>{\normalsize}c|>{\normalsize}c|>{\normalsize}c|>{\normalsize}c|}
\hline

Transmitter/ & \multicolumn{2}{>{\normalsize}c|}{Case I} & \multicolumn{2}{>{\normalsize}c|}{Case II} \\
\cline{2-5}
Receiver  &  Transmitted & Decoded & Transmitted & Decoded\\
 index & messages & messages & messages & messages \\
\hline
1 & $W_{11}$ & $\widehat{W}_{11}$,  $\widehat{W}_{12}$  & $W_{11}$ & $\widehat{W}_{11}$,  $\widehat{W}_{13}$\\
\hline
2 & $W_{12}$ & -  & $W_{22}$ & $\widehat{W}_{22}$\\
\hline
3 & $W_{33}$ & $\widehat{W}_{33}$  & $W_{13}$ & -  \\
\hline
\end{tabular}
}
\caption{Transmitted and Decoded messages for strategy $\mathcal{M}2$}
\label{table_strategy_M2}
\end{table}

{\bl The transmitted and decoded messages in strategy $\mathcal{M}2$ are illustrated in Table \ref{table_strategy_M2}. }
Here, we ask the following question: are there channel
conditions such that the sum-rate capacity is achieved by a two-user MAC at receiver 1 formed by 
transmitter 1 and either transmitter 2 (case I) or transmitter 3 (case II), while the interference from the other transmitter is treated as noise?
Observe that the other transmitter forms a point-to-point channel and is a source of interference for the two-user MAC.
We characterize the sum-rate capacity in the following theorem.

\begin{theorem}
{\bl For the $3 \times 3$ Gaussian many-to-one XC}, the sum-rate capacity is achieved by
strategy  $\mathcal{M}2$, where a two-user MAC is formed by transmitter 1 and either transmitter 2 or transmitter 3 at  
receiver 1, for the  following channel conditions, respectively
\begin{enumerate}
\item[(i)] $a^2 \;\; \ge  \;\; \displaystyle \frac{(1 + b^2 P_3)^2}{{1 - b^2}}, \quad \;\,b^2 \; <  \;  1$
\vspace{.3cm}
\item[(ii)] $b^2 \;\; \ge \; \; \displaystyle \frac{(1 + a^2 P_2)^2}{1 - a^2}, \quad \;\, a^2 \; < \; 1$.
\end{enumerate}
\label{thm_MAC_p2p}
\end{theorem}

\begin{IEEEproof}
We prove statement (i) below. {\bl This represents case I in Table \ref{table_strategy_M2}, where transmitters 1 
and 2 form a MAC at receiver 1 while interference from transmitter 3 is treated 
as noise. The proof for the second statement which corresponds to case II in Table \ref{table_strategy_M2} follows along similar lines.}

We use genie-aided bounding techniques to derive the optimality 
of strategy $\mathcal{M}2$. Specifically, we use the concept of 
\emph{useful genie} and \emph{smart genie} introduced 
in \cite{Veeravalli} to obtain the sum-rate capacity for strategy $\mathcal{M}2$.
Let a genie provide the following side information to receiver 1:
\begin{eqnarray}
s_1 & = & x_1 + a \, x_2 + \eta \,  z_1, \label{genie_side_inf}
\end{eqnarray}
where $z_1 \sim  \mathcal{N}(0, 1)$ and 
$\eta$ is a positive real number. We allow $z_1$ to be correlated 
to $n_1$ with correlation coefficient $\rho$.

A genie is said to be useful if it results in a genie-aided channel 
whose sum-rate capacity is achieved by Gaussian inputs, i.e., 
the sum-rate capacity of the genie-aided channel equals
$I({x}_{1G},\, {x}_{2G}\,;\, {y}_{1G}, {s}_{1G}) + I({x}_{3G} \,;\, {y}_{3G})$,
where $x_{iG} \sim \mathcal{N}(0, P_i)$, $y_{iG}$, $s_{1G}$ are $y_i$ 
and $s_1$ with $x_j = x_{jG}$, $\forall \,i, j$.

\begin{lemma}{(Useful Genie)}
The sum-rate capacity of the genie-aided channel with side 
information \eqref{genie_side_inf} given to receiver 1 is achieved
by using Gaussian inputs and by treating interference from 
transmitter 3 as noise at receiver 1, if the following conditions hold:
\begin{eqnarray}
\eta^2 \; \le \;  a^2, & \; & b^2 \; \le  \; 1 - \rho^2,\label{M2_useful_cond}
\end{eqnarray}
and the sum-rate of the genie-aided channel is 
bounded as 
\begin{eqnarray}
S & \le & I({x}_{1G},\, {x}_{2G}\,;\, {y}_{1G}, {s}_{1G}) + I({x}_{3G} \,;\, {y}_{3G}).
\label{genie_sum_cap_M2_3tx} 
\end{eqnarray}
\label{lem_useful_gen_M2}
\end{lemma}

\begin{IEEEproof}
The sum-rate of the genie-aided channel can be bounded as 
{\setlength{\arraycolsep}{5pt}
\begin{eqnarray}
n S & \le & H(W_{11}, W_{12}, W_{22}) + H(W_{13}, W_{33}) \nonumber \\
& = & I(W_{11}, W_{12}, W_{22} \,;\, {\bf y}_1^n, \,{\bf s}_1^n) + H(W_{11} \,|\, {\bf y}_1^n, \,{\bf s}_1^n)  +  H(W_{12} \,|\, {\bf y}_1^n, \,{\bf s}_1^n, {\bf x}_1^n)  
\nonumber \\
&& + \,H(W_{22}\, |\, {\bf y}_1^n, \,{\bf s}_1^n, {\bf x}_1^n, W_{12}) + I(W_{13}, W_{33} ; {\bf y}_3^n) \!+\! H(W_{13} | {\bf y}_3^n) \! +\!  H(W_{33} | {\bf y}_3^n, W_{13}) \nonumber \\
& \stackrel{(a)}{\le} & I({\bf x}_1^n, \, {\bf x}_2^n \, ;\, {\bf y}_1^n,  \,{\bf s}_1^n)  
+ {\bl H(W_{11} \,|\, {\bf y}_1^n)} +\, H(W_{12} \,|\, {\bf y}_1^n)  \nonumber \\
& &   + \,  H(W_{22} \,|\, {\bf s}_1^n,\, {\bf x}_1^n)  +  
I({\bf x}_3^n\, ;\, {\bf y}_3^n) + \,  H(W_{13} \,|\, {\bf y}_3^n) +  H(W_{33}\, |\, {\bf y}_3^n), \label{many_2_1_useful_gen_dec}
\end{eqnarray}
}where $(a)$ follows from the fact that removing conditioning cannot reduce the conditional  entropy.

We bound the term $H(W_{22} \,|\, {\bf s}_1^n,\, {\bf x}_1^n)$. If $\eta^2 \le a^2$, then
we have 
$ I(W_{22} \, ; \, {\bf s}_1^n \, | \, {\bf x}_1^n) \ge I(W_{22} \, ; \, {\bf y}_2^n)$. Thus, 
\begin{eqnarray}
H(W_{22} \,|\, {\bf s}_1^n, \, {\bf x}_1^n) & \le & H(W_{22} \,|\, {\bf y}_2^n) \nonumber\\
& \le & n \epsilon_n.\label{h_w22_s_1}
\end{eqnarray}
From Lemma \ref{lemma_dec_msg_rx_i}, we have $H(W_{13} \,|\, {\bf y}_3^n) \le n \epsilon_n$ when $b^2 \le 1$.
Using \eqref{many_2_1_fano_ineq} and  \eqref{h_w22_s_1}  in \eqref{many_2_1_useful_gen_dec}, we have 
{\setlength{\arraycolsep}{5pt}
\begin{eqnarray*}
n S & \le & I({\bf x}_1^n, {\bf x}_2^n \,;\, {\bf y}_1^n, \, {\bf s}_1^n) +  I({\bf x}_3^n \,;\, {\bf y}_3^n) + 5n \epsilon_n 
\label{m21_strat_M2_3tx_step1} \\
& = & I({\bf x}_1^n, {\bf x}_2^n ; {\bf s}_1^n) \!+ I({\bf x}_1^n, {\bf x}_2^n ; {\bf y}_1^n \, | \, {\bf s}_1^n) \!+ I({\bf x}_3^n ; {\bf y}_3^n) + 5n \epsilon_n\\
& = &  h({\bf s}_1^n) - h({\bf s}_1^n \, | \, {\bf x}_1^n, {\bf x}_2^n) + h({\bf y}_1^n \, | \, {\bf s}_1^n) \nonumber \\
& & -\, h({\bf y}_1^n \, | \, {\bf s}_1^n, {\bf x}_1^n, {\bf x}_2^n) + h({\bf y}_3^n) 
- h({\bf y}_3^n \, | \, {\bf x}_3^n)+ 5n \epsilon_n \nonumber \\
&  = & h({\bf s}_1^n) - h(\eta \, {\bf z}_1^n) +  h({\bf y}_1^n \, | \, {\bf s}_1^n) 
 - h(b\,  {\bf x}_3^n + {\bf n}_1^n \, | \, {\bf z}_1^n)  
+\, h({\bf y}_3^n) - h({\bf n}_3^n) + 5n \epsilon_n\nonumber \\
& \stackrel{(b)}{\le} & n h({ s}_{1G}) - n h(\eta \,  {z}_1) + {\bl n  h({ y}_{1G} \, | \, {s}_{1G}) } \nonumber \\
& & -\, h(b\,  {\bf x}_3^n + \tilde{\bf n}_1^n)  + h({\bf x}_3^n + {\bf n}_3^n) - n h({n}_3) + 5n \epsilon_n\nonumber \\
& \stackrel{(c)}{\le} & n h({ s}_{1G}) - n h(\eta \,  {z}_1) +  {\bl n h({ y}_{1G} \, | \, {s}_{1G}) } \nonumber \\
& &  +  \, n h({ x}_{3G} + { n}_3) - nh(b \, {x}_{3G} + \tilde{n}_1)  -  n h({n}_3) + 5n \epsilon_n \nonumber \\
& = &  n \,I({x}_{1G}, {x}_{2G} \,;\, {y}_{1G}, \, {s}_{1G}) +  n \,I({x}_{3G} \,;\, {y}_{3G})  + 5n \epsilon_n,
\end{eqnarray*}
}where $\tilde{n_1}  \sim \mathcal{N}(0, 1 - \rho^2)$, $(b)$ follows since  Gaussian inputs maximize differential entropy for a given covariance constraint and from the application of Lemmas 1 and 6 in \cite{Veeravalli}, 
$(c)$ follows from applying Lemma 1 in \cite{Motahari} (which is a special case of the extremal inequality considered in \cite{Liu2}) 
to the term $h({\bf x}_3^n + {\bf n}_3^n) -\, h(b\,  {\bf x}_3^n + \tilde{\bf n}_1^n)$, 
and using the condition $b^2 \; \le  \; 1 - \rho^2$. \end{IEEEproof}

Next, we show that the genie is smart. A smart genie is one which does not 
improve the sum-rate when Gaussian inputs are used, i.e., 
$I({x}_{1G},\, {x}_{2G}\, ; \, {y}_{1G}, {s}_{1G}) = 
I({x}_{1G},\, {x}_{2G} \, ; \, {y}_{1G})$.

\begin{lemma}{(Smart Genie)}
If Gaussian inputs are used, and interference is treated as noise, 
then, under the condition
\begin{eqnarray}
\eta \rho = 1\; + \; b^2 P_3,  \label{many_2_1_smart_cond}
\end{eqnarray}
 the genie does not increase the sum rate, i.e.,
\begin{eqnarray}
I({ x}_{1G}, \,{x}_{2G} \, ; \, { y}_{1G}, \, {s}_{1G}) = I({x}_{1G}, \, {x}_{2G} \, ; \, {y}_{1G}).
\label{smart_genie_sum_rate}
\end{eqnarray}
\label{lem_smart_gen}
\end{lemma}

\begin{IEEEproof}
Note that  
\begin{eqnarray*}
I({x}_{1G}, \,{x}_{2G} \, ;\,  { y}_{1G}, {s}_{1G}) 
& = &  I({x}_{1G}, \,{x}_{2G} \,; \, { y}_{1G}) + 
I({x}_{1G},\, {x}_{2G} \, ; \, { s}_{1G} \, |\, { y}_{1G}).
\end{eqnarray*}
The second term on the right hand side can be expanded as 
\begin{eqnarray*}
I({ x}_{1G} \,;\, { s}_{1G} \,|\, { y}_{1G}) + I({x}_{2G} \,;\, {s}_{1G} \, | \, { y}_{1G}, \, { x}_{1G}).
\end{eqnarray*}
Consider
\begin{eqnarray*}
I({ x}_{1G} ; { s}_{1G} \, |\, { y}_{1G}) & = &I({x}_{1G} \,;\,  x_{1G} + a \, x_{2G} + \eta z_1 \, | \, 
x_{1G} + a\,x_{2G} + b\,x_{3G}+ n_1).
\end{eqnarray*}
From Lemma 8 in \cite{Veeravalli}, if $x$, $n$, $z$ are Gaussian with $x$ 
being independent of the two zero-mean random variables $n$, $z$, then 
$I(x \,;\, x + z \,|\, x + n) = 0$, iff $\mathbb{E}(z\, n) = \mathbb{E}(n^2)$.
Thus, $I({ x}_{1G} ; { s}_{1G} \, |\, { y}_{1G})$ becomes zero if 
$a^2  P_2 + \eta \, \rho = 1 + a^2  P_2 + b^2 P_3$
which reduces to \eqref{many_2_1_smart_cond}. Now, consider
\begin{eqnarray*}
I({x}_{2G} \, ; \, {s}_{1G} \, | \, { y}_{1G}, \, { x}_{1G}) 
& = & \; I({x}_{2G} \, ; \, a \, x_{2G} + \eta \, z_1 \, | \, 
a\, x_{2G} + b \, x_{3G} + n_1) \nonumber \\
& \stackrel{(d)}{=} & 0. 
\end{eqnarray*}
where $(d)$ follows from \cite[Lemma 8]{Veeravalli} and \eqref{many_2_1_smart_cond}. \end{IEEEproof}

Combining conditions \eqref{M2_useful_cond} and \eqref{many_2_1_smart_cond}, we have 
{\bl\begin{eqnarray}
a^2 \;\; \ge\;\; \frac{(1 + b^2 P_3)^2}{\rho^2}  & ; & b^2 \; \le \; {1 - \rho^2}.
\label{m21_comb_conditions_1}
\end{eqnarray}}
For a fixed value of $b$, we have the constraint $\rho^2 \le {1 - b^2}$. 
Note that choosing $\rho^2 = {1 - b^2}$ results in the best bound for $a^2$. {\bl From \eqref{many_2_1_smart_cond}, we infer that $\rho > 0$, and using  \eqref{M2_useful_cond}, this implies that $b^2 < 1$.}
Thus, \eqref{m21_comb_conditions_1} can be rewritten as statement (i) 
in Theorem \ref{thm_MAC_p2p}. \end{IEEEproof}

\subsection{Gap from optimality of Strategy $\small \mathcal{M}3$}

\begin{table}[t]
\centering
{
\renewcommand{\arraystretch}{1.2}
\renewcommand{\tabcolsep}{.2cm}
\begin{tabular}{|>{\normalsize}c|>{\normalsize}c|>{\normalsize}c|}
\hline
 Transmitter/  &  Transmitted & Decoded  \\
Receiver index & messages & messages \\
\hline
1 & $W_{11}$ & $\widehat{W}_{11}$, $\widehat{W}_{12}$, $\widehat{W}_{13}$ \\
\hline
2 & $W_{12}$ & -  \\
\hline
3 & $W_{13}$ &  -  \\
\hline
\end{tabular}
}
\caption{Transmitted and Decoded messages for strategy $\mathcal{M}3$}
\label{table_strategy_M3}
\end{table}

{\bl The transmitted and decoded messages in strategy $\mathcal{M}3$ are illustrated in Table \ref{table_strategy_M3}.} 
In strategy $ \mathcal{M}3$, all transmitters form a MAC at receiver 1. 
We derive a sum-rate outer bound to the
 many-to-one XC and characterize the gap between the outer bound and the 
achievable sum-rate of strategy $\mathcal{M}3$.

\begin{theorem}
{\bl For the $3 \times 3$ Gaussian many-to-one XC, when strategy $\mathcal{M}3$ is employed}, if 
\begin{eqnarray}
a^2 \,\, \ge \,\, \displaystyle \frac{(1 + b^2 P_3)^2}{\rho^2}  & \text{and} & \;\, b^2 \,\, \ge\,\, 1,   
\label{many_2_one_strt2_sub_reg}
\end{eqnarray}
then the gap between the sum-rate outer bound and the sum-rate of 
 strategy $\small \mathcal{M}3$ is given by
\begin{eqnarray}
0.5 \log \left( \frac{1 - ({1 + b^2 P_3})^{-1}\rho^2 }{1 - \rho^2} \right),
\end{eqnarray}
where $\rho$ denotes a constant with {\bl $\rho \in [-1, 1]$}.
\label{thm_MAC_rx1}
\end{theorem}

\begin{IEEEproof}
We use genie-aided techniques to derive the sum-rate outer bound. 
Let a genie provide the side information given in \eqref{genie_side_inf} 
to receiver 1.
We prove below that the genie is useful. 

\begin{lemma}{(Useful Genie)}
The sum-rate capacity of the genie-aided channel with side 
information \eqref{genie_side_inf} given to receiver 1 is achieved
by using Gaussian inputs when all transmitters transmit to receiver 1, 
if the following conditions hold:
\begin{eqnarray}
\eta^2 \;\; \le \;\;  a^2, & \; & b^2 \;\; \ge  \;\; 1, \label{useful_cond}
\end{eqnarray}
and the sum-rate of the genie-aided channel is 
bounded as 
\begin{eqnarray}
S & \le & I({x}_{1G},\, {x}_{2G}, \, {x}_{3G} \, ;\, {y}_{1G}, {s}_{1G}).
\label{genie_sum_cap_M3_3tx}
\end{eqnarray}
\label{m21_useful_genie_M3}
\end{lemma}

\begin{IEEEproof}
The sum-rate $S$ of the genie-aided channel is bounded as
{\setlength{\arraycolsep}{5pt}
\begin{eqnarray}
n S & \le & H(W_{11}, W_{12}, W_{13}, W_{22},W_{33}) \nonumber \\
&  = & {\bl I(W_{11}, W_{12}, W_{13}, W_{22},W_{33} \, ; \, {\bf y}_1^n, {\bf s}_1^n)  + H(W_{11}, W_{12}, W_{13}, W_{22},W_{33} \, | \, {\bf y}_1^n, {\bf s}_1^n) } \nonumber \\
& = & {\bl I(W_{11}, W_{12}, W_{13}, W_{22},W_{33} \, ; \, {\bf y}_1^n, {\bf s}_1^n) }
+  H(W_{11} \,| \,{\bf y}_1^n, {\bf s}_1^n) 
+  H(W_{12} \,|\, {\bf y}_1^n,  {\bf s}_1^n, {\bf x}_1^n)  \nonumber\\
&&+\, H(W_{22} | {\bf y}_1^n,  {\bf s}_1^n, {\bf x}_1^n, W_{12}) 
  + H(W_{13} | {\bf y}_1^n, {\bf s}_1^n, {\bf x}_1^n,  {\bf x}^n_2)  
  +  H(W_{33} | {\bf y}_1^n,  {\bf s}_1^n, {\bf x}_1^n,  {\bf x}^n_2, W_{13}) \nonumber \\
& \le & I({\bf x}_1^n, {\bf x}_2^n, {\bf x}_3^n ; {\bf y}_1^n,  {\bf s}_1^n)  
+ H(W_{11} | {\bf y}_1^n) + H(W_{12} | {\bf y}_1^n) 
+ H(W_{22} | {\bf s}_1^n, {\bf x}_1^n) \!+\! H(W_{13} | {\bf y}_1^n) \nonumber \\
& & +  H(W_{33} | {\bf y}_1^n, {\bf x}_1^n,  {\bf x}^n_2). \label{m21_mac_rx1_dec} 
\end{eqnarray}
}

We bound the term $H(W_{33} \,|\, {\bf y}_1^n, {\bf x}_1^n, {\bf x}_2^n)$. 
If $b^2 \ge 1$, then $ I(W_{33} ; {\bf y}_1^n \,|\, {\bf x}_1^n, {\bf x}_2^n) \ge 
I(W_{33} \,;\, {\bf y}_3^n) $. Therefore, 
\begin{eqnarray}
H(W_{33} \,|\, {\bf y}_1^n, \, {\bf x}_1^n, {\bf x}_2^n) & \le & H(W_{33} \,|\, {\bf y}_3^n)  \nonumber \\
& \le & n \epsilon_n.
\label{H_W_33_y_1_x1_x2}
\end{eqnarray}
Note that the term $H(W_{22} \,|\, {\bf s}_1^n, {\bf x}_1^n)$ is again bounded as in \eqref{h_w22_s_1} if $\eta^2 \; \le \;  a^2$.
Using \eqref{many_2_1_fano_ineq}, \eqref{h_w22_s_1}, and \eqref{H_W_33_y_1_x1_x2}
in \eqref{m21_mac_rx1_dec}, we have 
\begin{eqnarray}
nS & \le  & I({\bf x}_1^n, {\bf x}_2^n, {\bf x}_3^n ; {\bf y}_1^n,  {\bf s}_1^n)  + 5 n \epsilon_n \nonumber \\
& = & I({\bf x}_1^n, {\bf x}_2^n, {\bf x}_3^n ; {\bf y}_1^n) + 
I({\bf x}_1^n, {\bf x}_2^n, {\bf x}_3^n ; {\bf s}_1^n \, | \, {\bf y}_1^n)+ 5 n \epsilon_n \nonumber \\
& \stackrel{(a)}{\le} &  n I({x}_{1G}, {x}_{2G}, {x}_{3G} ; {y}_{1G}) + h({\bf s}_1^n \, | \, {\bf y}_1^n)
- h({\bf s}_1^n \, | \, {\bf y}_1^n, {\bf x}_1^n, {\bf x}_2^n, {\bf x}_3^n) + 
5 \epsilon_n \nonumber \\
& \stackrel{(b)}{\le} & n I({x}_{1G}, {x}_{2G}, {x}_{3G} ; {y}_{1G}) 
+ n h({s}_{1G} \, | \, {y}_{1G})   - nh(\eta z_1 \, | \, n_1) + 5 \epsilon_n \nonumber \\
& = &  n I({x}_{1G},\, {x}_{2G}, \, {x}_{3G} \, ;\, {y}_{1G}, {s}_{1G}) + 
5 \epsilon_n, \nonumber
\end{eqnarray}
where $(a)$ follows from the optimality of Gaussian inputs for Gaussian MAC, 
$(b)$ follows from Lemma 1 in \cite{Veeravalli}.
Here, ${y}_{1G}$ denotes $y_1$ with $x_i$ being Gaussian distributed, 
i.e., ${y}_{1G} = {x}_{1G} + a {x}_{2G} + b {x}_{3G} + { n}_1$.
As $n \rightarrow \infty$, $\epsilon_n \rightarrow 0$ and we get the desired bound. \end{IEEEproof}

Unlike in the case of strategy  $\small \mathcal{M}2$, here the genie does 
in fact increase the sum-rate and hence is not smart. 
However, we can choose the parameters $\rho$ 
and $\eta$ to get a good sum-rate outer bound as follows. 
Consider 
\begin{eqnarray}
I({x}_{1G},\, {x}_{2G}, \, {x}_{3G} \, ;\, {y}_{1G}, {s}_{1G}) 
& = & \; I({x}_{1G}, {x}_{2G}, {x}_{3G} ; {y}_{1G})  + 
I({x}_{1G}, {x}_{2G}, {x}_{3G} ; {s}_{1G} \, | \, {y}_{1G}). \nonumber 
\end{eqnarray}
The second term on the right hand side can be expanded as 
\begin{eqnarray}
I({x}_{1G},\, {x}_{2G} ; { s}_{1G} \, |\, { y}_{1G}) + 
I({x}_{3G} \, ; \, { s}_{1G} \, |\, { y}_{1G}, {x}_{1G},\, {x}_{2G}).
\end{eqnarray}

\begin{table}[t]
\centering
{
\renewcommand{\arraystretch}{2.5}
\renewcommand{\tabcolsep}{.4cm}
\begin{tabular}{|>{\normalsize}c|>{\normalsize}c|>{\normalsize}c|}
\hline
 Strategy &  {Channel conditions} & Gap from Outer-bound\\
\hline
\normalsize {$\normalsize \mathcal{M}1$ } & 
$a^2 + b^2 \le 1$ & 0\\
\hline
\normalsize {$\normalsize \mathcal{M}2$ } & (i)  $a^2 \ge \displaystyle \frac{(1 + b^2 P_3)^2}{1 - b^2}, \;\, {\bl b^2 < 1}$ \normalsize   & 0 
\\
\cline{2-3}
&  (ii) $b^2 \ge \displaystyle \frac{(1 + a^2 P_2)^2}{1 - a^2}, \;\, {\bl a^2 < 1} $ & 0 \\
\hline
\normalsize {$\normalsize \mathcal{M}3$ } &  (i) 
$a^2 \, \ge \displaystyle \frac{(1 + b^2 P_3)^2}{\rho^2} , \;\, b^2 \ge 1$ & 
$0.5\log \Bigg[\displaystyle \frac{1 - \displaystyle\frac{\rho^2}{1 + b^2\, P_3}}{1 - \rho^2} \Bigg]$   \rule{0pt}{1cm}
\\
\cline{2-3}
&(ii)  $b^2 \, \ge \displaystyle \frac{(1 + a^2 P_2)^2}{\rho^2} , \;\, a^2 \ge 1$ & 
$0.5 \log \Bigg[\displaystyle \frac{1 - \displaystyle\frac{\rho^2}{1 + a^2\, P_2}}{1 - \rho^2} \Bigg]$ \rule{0pt}{1cm}  \\
\hline
\end{tabular}
}
\caption{Summary of results for many-to-one X channel}
\label{table_many21_sum_results}
\end{table}

In the proof of Lemma \ref{lem_smart_gen}, we showed that by choosing
$\eta \, \rho = 1 +b^2 P_3$, we can make $I({x}_{1G},\, {x}_{2G} ; { s}_{1G} \, |\, { y}_{1G}) = 0$.  Now, consider 
\begin{eqnarray}
I({x}_{3G} ; {s}_{1G} \, | \, { y}_{1G}, \, { x}_{1G}, { x}_{2G}) & = & I({x}_{3G} ; \eta \, z_1 \, | \, 
b\, x_{3G} + n_1) \nonumber \\
&  = &  h(\eta \, z_1 \, | \, b\, x_{3G} + n_1) - h(\eta \, z_1 \, | \, n_1)  \nonumber \\
& \stackrel{(c)}{=} &   h(\eta \, z_1 \, | \, b\, x_{3G} + n_1) - h(\eta \, \tilde{z}_1)  \nonumber \\
& =  &  0.5\log \left( \frac{\eta^2 (1 + b^2 \, P_3) - \eta^2 \rho^2}{(1 + b^2 P_3)\eta^2 (1 - \rho^2)} \right)  \nonumber \\
& = &  0.5 \log \left( \frac{1 - ({1 + b^2 P_3})^{-1}\rho^2 }{1 - \rho^2} \right)\!, \label{sum_rate_gap} 
\end{eqnarray}
where $\tilde{z}_1 \sim \mathcal{N}(0, 1 - \rho^2)$ and $(c)$ follows from 
\cite[Lemma 6]{Veeravalli}. Note that \eqref{sum_rate_gap} represents the gap between 
the sum-rate outer bound and the sum-rate of strategy  $\small \mathcal{M}3$.
Combining condition \eqref{useful_cond} with $\eta \, \rho = 1 +b^2 P_3$, we get \eqref{many_2_one_strt2_sub_reg}. \end{IEEEproof}

Due to the underlying symmetry in the MAC at receiver 1, a result corresponding 
to Theorem \ref{thm_MAC_rx1} with the channel coefficients $a$, $b$ and power levels 
$P_2$, $P_3$ interchanged is also true and further can be proved along 
similar lines. The results of this section are succinctly summarized in Table \ref{table_many21_sum_results}.

\subsection{Recovering known results for the Z channel}

We specialize the results in this section to the Z channel. The Z channel is 
obtained from the many-to-one X channel by retaining only the 
first two transmitters and removing the rest \cite{Liu, Chong}. In the
$3 \times 3$ many-to-one XC shown in Fig.
\ref{many_2_1_3tx_std_form}, this is equivalent
to setting $b = 0$,  and considering the outputs at the
first two receivers alone. In this case, Theorem \ref{thm_low_interf} reduces
to the channel condition $a^2 \le 1$, which is identical to 
that obtained in \cite{Liu} for the low-interference regime. 
Theorem \ref{thm_MAC_p2p} reduces to the condition $a^2 \ge 1$, which is
same as that obtained in \cite{Chong} for the MAC sum-rate at receiver 1 to be
the sum-rate capacity of the Z channel.

\section{Extension to the $K \times K$ Many-to-One X Channel}
\label{sec_exten_K_tx}

Since the results for the $K \times K$ many-to-one XC follow more or 
less along similar lines as the  $3 \times 3$ case, we state the 
results along with a brief outline of the proof for each strategy, with 
additional details provided in places where the proofs differ.

\subsection{Conditions for the sum-rate optimality of strategies 
$\mathcal{M}1$, $\mathcal{M}2$ and $\mathcal{M}3$}
\label{subsec_cond_sum_rate_opt_m21_xc}
The optimality of strategy $\small \mathcal{M}1$ follows using similar arguments 
as in Theorem \ref{thm_low_interf}, under the condition $\sum_{i\,=\,2}^{K} h_i^2 \le 1$.
This condition arises from the use of Lemma \ref{lemma_li}, as in inequality $(b)$ 
of Theorem \ref{thm_low_interf}. To avoid repeating the details, we omit the proof.

Next, we consider the optimality of strategy $\small \mathcal{M}2$.
Here, we are interested in a region where the sum-rate capacity is achieved by a 
two-user MAC at receiver 1 formed by transmitter 1 and transmitter $k$, 
$k = 2, \ldots, K$, while the interference from the other transmitters is 
treated as noise. {\bl In strategy $\small \mathcal{M}2$, the transmitted messages are $W_{ii}$ at transmitter $i$, $i \ne k$, and 
$W_{1k}$ at transmitter $k$. The decoded messages are $(\widehat{W}_{11}, \widehat{W}_{1k})$ at receiver 1, and $\widehat{W}_{jj}$ at receiver $j$, $j \ne (1, k)$.}
We characterize the sum-rate capacity in the following theorem.

\begin{theorem}
{\bl For the $K \times K$ Gaussian many-to-one XC}, 
the sum-rate capacity is achieved by the two-user MAC formed by transmitter 1
and transmitter $k$ to receiver 1, for the following channel conditions
\begin{eqnarray}
h_k^2 & \ge & \frac{\Big(1 + \sum_{\substack{j=2,\,j\ne k}}^{K} h_j^2
P_j\Big)^2}
{1-\sum_{\substack{j=2,\,j\ne k}}^{K}h_j^2}, 
\nonumber \\
{\bl \sum_{\substack{j=2,\,j\ne k}}^{K}h_j^2} & < & 1. 
\label{m21_sub_reg_Ktx_M2}
\end{eqnarray}
\label{thm_MAC_p2p_Ktx}
\end{theorem}

\begin{IEEEproof}
Let a genie provide the following side information to receiver 1:
\begin{eqnarray}
s_k & = & x_1 + h_k \, x_k + \eta_k \,  z_k, \label{genie_side_inf_Ktx_M2}
\end{eqnarray}
where $z_k \sim  \mathcal{N}(0, 1)$ and 
$\eta_k$ is a positive real number. We allow $z_k$ to be correlated 
to $n_1$ with correlation coefficient $\rho_k$.

\begin{lemma}{(Useful Genie)}
The sum-rate capacity of the genie-aided channel with side 
information \eqref{genie_side_inf_Ktx_M2} given to receiver 1 is achieved
by using Gaussian inputs and by treating interference as noise
at receiver 1, if the following conditions hold:
\begin{eqnarray}
\eta_k^2 \; \le \; h_k^2, & \; & \sum_{\substack{j=2,\,j\ne k}}^{K} h_j^2 \; \le \; 1 -
\rho_k^2.\label{many_2_1_Ktx_useful_cond_M2}
\end{eqnarray}

\end{lemma}

\begin{IEEEproof}
{\bl
The sum-rate of the genie-aided channel can be bounded as 
{\setlength{\arraycolsep}{5pt}
\begin{eqnarray}
n S & \le & H(W_{11}, W_{1k}, W_{kk}) + \sum_{j=2, j \ne k}^K H(W_{1j}, W_{jj}) \nonumber \\
& = & I(W_{11}, W_{1k}, W_{kk} \,;\, {\bf y}_1^n, \,{\bf s}_k^n) + H(W_{11} \,|\, {\bf y}_1^n, \,{\bf s}_k^n)  +  H(W_{1k} \,|\, {\bf y}_1^n, \,{\bf s}_k^n, {\bf x}_1^n)  
\nonumber \\
&& + \,H(W_{kk}\, |\, {\bf y}_1^n, \,{\bf s}_k^n, {\bf x}_1^n, W_{1k}) 
+  \sum_{j=2, j \ne k}^K I(W_{1j}, W_{jj} \, ; \, {\bf y}_j^n) +   \sum_{j=2, j \ne k}^K  \big[ H(W_{1j} \, | \, {\bf y}_j^n) \nonumber \\  
& & + \, H(W_{jj} \, | \, {\bf y}_j^n, \, W_{1j})  \big]\nonumber \\
& \stackrel{(a)}{\le} & I({\bf x}_1^n, \, {\bf x}_k^n \, ;\, {\bf y}_1^n,  \,{\bf s}_k^n)  
+ {\bl H(W_{11} \,|\, {\bf y}_1^n)} +\, H(W_{1k} \,|\, {\bf y}_1^n)  \nonumber \\
& &   + \,  H(W_{kk} \,|\, {\bf s}_k^n,\, {\bf x}_1^n)  +  
 \sum_{j=2, j \ne k}^K I({\bf x}_j^n\, ;\, {\bf y}_j^n) +\!\!    \sum_{j=2, j \ne k}^K \!\!\big[ H(W_{1j} \,|\, {\bf y}_j^n) +  H(W_{jj}\, |\, {\bf y}_j^n) \big], \label{many_2_1_useful_gen_dec_Ktx}
\end{eqnarray}
}where $(a)$ follows from the fact that removing conditioning cannot reduce the conditional  entropy.

As in Lemma \ref{lem_useful_gen_M2}, if  $\eta^2 \le a^2$, we have 
\begin{eqnarray}
H(W_{kk} \,|\, {\bf s}_k^n, \, {\bf x}_1^n) & \le & H(W_{kk} \,|\, {\bf y}_k^n) \; \le \; n \epsilon_n. \label{h_w22_s_1_Ktx}
\end{eqnarray}
From Lemma \ref{lemma_dec_msg_rx_i}, if $h_j^2 \le 1$, we have $H(W_{1j} \,|\, {\bf y}_j^n) \le n \epsilon_n$. Using this along with \eqref{many_2_1_fano_ineq} and 
\eqref{h_w22_s_1_Ktx} in \eqref{many_2_1_useful_gen_dec_Ktx}, we have}
\begin{eqnarray}
n S & \le & I({\bf x}_1^n, {\bf x}_k^n \,;\, {\bf y}_1^n, \, {\bf s}_k^n) +
\sum_{j=2, \,j\ne k}^{K} I({\bf x}_j^n \,;\, {\bf y}_j^n) + (2K-1)n \epsilon_n \nonumber
\\
& = & I({\bf x}_1^n, {\bf x}_k^n \,;\, {\bf s}_k^n)
+ I({\bf x}_1^n, {\bf x}_k^n \,;\, {\bf y}_1^n \, | \, {\bf s}_k^n)
+ \sum_{j=2, \,j\ne k}^{K} I({\bf x}_j^n \,;\, {\bf y}_j^n) + (2K-1)n \epsilon_n \nonumber \\
& = & h({\bf s}_k^n) - h({\bf s}_k^n \,|\, {\bf x}_1^n, {\bf x}_k^n) + h({\bf y}_1^n \,
|\, {\bf s}_k^n) - h({\bf y}_1^n \, |\, {\bf s}_k^n, {\bf x}_1^n, {\bf x}_k^n)
\nonumber \\
&& + \sum_{j=2, \,j\ne k}^{K} \big[h({\bf y}_j^n) - h({\bf y}_j^n \, | \, {\bf x}_j^n) \big]+
(2K-1)n \epsilon_n \nonumber \\
& = & h({\bf s}_k^n) - h(\eta_k \, {\bf z}_k^n) + h({\bf y}_1^n \, | \, {\bf
s}_k^n)
- h\Big(\sum_{j=2, \,j\ne k}^{K} h_j\, {\bf x}_j^n + {\bf n}_1^n \, | \, {\bf z}_k^n\Big)
\nonumber \\
& &+\, \sum_{j=2, \,j\ne k}^{K} \big[h({\bf y}_j^n) - h({\bf n}_j^n)\big] + (2K-1)n
\epsilon_n\nonumber \\
& \stackrel{(b)}{\le} & n h({ s}_{kG}) - n h(\eta_k \, {z}_k) + n h({ y}_{1G} \,
| \, {s}_{kG}) - h\Big(\sum_{j=2, \,j\ne k}^{K} h_j\, {\bf x}_j^n + \tilde{\bf n}_1^n\Big)
\nonumber \\
& & + \sum_{j=2, \,j\ne k}^{K} \big[ h({\bf x}_j^n + {\bf n}_j^n) - n h({n}_j)\big] +
(2K-1) \epsilon_n\nonumber \\
& \stackrel{(c)}{\le} & n h({ s}_{kG}) - n h(\eta_k \, {z}_k) + n h({ y}_{1G} \,
| \, {s}_{kG}) + \sum_{j=2, \,j\ne k}^{K} n h({x}_{jG} + {n}_j)
\nonumber \\
& &  - n h\Big(\sum_{j=2, \,j\ne k}^{K} h_j\,  {x}_{jG} + \tilde{n}_1\Big) 
 -  \sum_{j=2, \,j\ne k}^{K} n h({n}_j) +(2K-1) \epsilon_n \nonumber \\
& = & n \,I({x}_{1G}, {x}_{kG} \,;\, {y}_{1G}, \, {s}_{kG}) + \sum_{j=2, \,j\ne k}^{K} n
\,I({x}_{jG} \,;\, {y}_{jG}) + (2K-1) \epsilon_n,
\label{m21_useful_gen_sum_rate_M2}
\end{eqnarray}
where $\tilde{n_1} \sim \mathcal{N}(0, 1 - \rho_k^2)$, $(b)$ follows since
Gaussian inputs maximize
differential entropy for a given covariance constraint and from the application
of Lemma 1 and Lemma 6 in \cite{Veeravalli},
$(c)$ follows from applying Lemma \ref{lemma_li}
to the term $\sum_{j=2, \,j\ne k}^{K} h({\bf x}_j^n + {\bf n}_j^n) -\, 
h\big(\sum_{j=2, \,j\ne k}^{K} h_j\, {\bf x}_j^n + \tilde{\bf n}_1^n\big)$, and using the
condition $\sum_{j=2, \,j\ne k}^{K} h_j^2 \; \le \; 1 - \rho_k^2$. \end{IEEEproof}

Using similar arguments as in Lemma \ref{lem_smart_gen}, the genie is
smart if 
\begin{eqnarray}
\eta_k \rho_k = 1\; + \; \sum_{\substack{j=2,\,j\ne k}}^K h_j^2  P_j,  \label{many_2_1__Kx_smart_cond}
\end{eqnarray}
which ensures that the genie does not increase the sum rate, i.e., $I({ x}_{1G}, \,{x}_{kG} \, ; \, { y}_{1G}, \, {s}_{kG}) = I({x}_{1G}, \,
{x}_{kG} \, ; \, {y}_{1G})$.
As before, the conditions \eqref{many_2_1_Ktx_useful_cond_M2} and 
\eqref{many_2_1__Kx_smart_cond} can be combined  to get \eqref{m21_sub_reg_Ktx_M2}. \end{IEEEproof}

{\bl 
The characterization of the optimality of strategies where more than two transmitters 
form a MAC at receiver 1 can theoretically be obtained using similar techniques as in 
Theorem \ref{thm_MAC_rx1} and Theorem \ref{thm_MAC_p2p_Ktx}. However, we note that as in Theorem \ref{thm_MAC_rx1}, the genie is no longer smart and results in a sum-rate outer bound for the $K \times K$ many-to-one XC. As before, the gap between this outer bound and achievable sum-rate of the strategy can be characterized. However, we defer this to a future work as the characterization of the gap from the outer bound is decidedly more complicated.}

\subsection{A region in which the many-to-one XC can be operated 
as a many-to-one IC }

We identify a region in which the many-to-one XC can be operated as a 
many-to-one IC without loss of sum-rate. To accomplish this, we need to show 
that the absence of cross messages does not lead to a decrease in the sum-rate. 
We have the following result. 

\begin{theorem}
The $K \times K$ many-to-one XC can be operated as a $K$-user 
many-to-one IC without loss of sum-rate in the following sub-region
\begin{eqnarray}
h_i^2 & \le & 1,  \qquad i = 2, \ldots, K. \label{sub_reg_oper_xc_ic_Ktx}
\end{eqnarray}
\label{thm_oper_m21_xc_as_ic_Ktx}
\end{theorem}

\begin{IEEEproof}
Let $h_i^2 \le 1$,  $i = 2, \ldots, K$.
The sum-rate can be bounded as follows:
\begin{eqnarray}
n S & = & H(W_{11}) + \sum_{k=2}^{K} H(W_{1k}, W_{kk}) \nonumber \\
& = & I(W_{11} \, ; \, {\bf y}_1^n) + \sum_{k=2}^{K} I(W_{1k}, W_{kk} \, ; \, {\bf y}_k^n) + {\bl H(W_{11} \, | \, {\bf y}_1^n) +  \sum_{k=2}^{K} H(W_{1k}, W_{kk} \, |\,  {\bf y}_k^n) } \nonumber \\
& \le & \sum_{k=1}^{K} I({\bf x}_k^n \, ;\, {\bf y}_k^n) + (2K-1) n \epsilon_n, \label{sum_rate_eq_m21_IC_Ktx} 
\end{eqnarray}
{\bl where \eqref{sum_rate_eq_m21_IC_Ktx} follows from \eqref{many_2_1_fano_ineq_Ktx} and the application of Lemma \ref{lemma_dec_msg_rx_i} when $h_i^2 \le 1,  i = 2, \ldots, K$.}
We note that \eqref{sum_rate_eq_m21_IC_Ktx} 
is in fact the sum-rate of the corresponding $K \times K$ many-to-one IC.
From \eqref{sum_rate_eq_m21_IC_Ktx}, it is clear that we can set $W_{1k} = \phi$, $k = 2, \ldots, K$ (without loss of sum-rate). 
Thus, we have shown that the absence of cross messages does not diminish the 
sum-rate when  $h_i^2 \le 1$,  $i = 2, \ldots, K$.  \end{IEEEproof}

\subsection{Conditions for sum-rate of strategy of $\mathcal{M}1$
to be within $K/2-1$ bits from sum-rate capacity}

In the following theorem, we show that in sub-region \eqref{sub_reg_oper_xc_ic_Ktx}, 
strategy $\mathcal{M}1$, i.e., using Gaussian codebooks and treating interference as noise, 
can achieve a sum-rate to within $K/2-1$ bits  from  the sum-rate capacity of the Gaussian many-to-one XC. 

\begin{theorem}
{\bl For the $K \times K$ Gaussian many-to-one XC},
in sub-region \eqref{sub_reg_oper_xc_ic_Ktx}, the rate point achieved by 
strategy $\mathcal{M}1$, i.e., using Gaussian codebooks and treating interference 
as noise is within $K/2-1$ bits from  the sum-rate capacity of Gaussian 
many-to-one XC.
\label{thm_m21_xc_char_1_bit_per_user}
\end{theorem}

\begin{IEEEproof}
Assume $h_i^2 \le 1$, $i = 2, \ldots, K$, i.e., sub-region \eqref{sub_reg_oper_xc_ic_Ktx} 
is true.
Let a genie provide the following side-information to receiver $i$, 
$ i = 2, \ldots, K-1$
\begin{eqnarray}
s_i & = & \sum_{j\,=\,i}^K h_j \, x_j + n_1.
\label{gen_sig_char_sum_rate_xc_ic_Ktx}
\end{eqnarray}
Using Theorem \ref{thm_oper_m21_xc_as_ic_Ktx},  receiver $i$ is able to decode $(W_{ii}, W_{1i})$ 
in sub-region \eqref{sub_reg_oper_xc_ic_Ktx}, with or without  the genie signals.
Hence, the sum-rate of the genie-aided channel is bounded as follows:
\begin{eqnarray}
n S & \le &  I({\bf x}_1^n \,;\, {\bf y}_1^n) 
+ \sum_{i=2}^{K-1} I({\bf x}_i^n \, ; \, {\bf y}_i^n, \,  {\bf s}_i^n)
+ I({\bf x}_K^n \,;\, {\bf y}_K^n) + (2K-1) n \epsilon_n \label{sum_rate_eq_m21_IC_gen_aid_Ktx} \\
& = & h({\bf y}_1^n)  - h({\bf y}_1^n\, | \, {\bf x}_1^n) + 
 \sum_{i=2}^{K-1} \big[
 I({\bf x}_i^n \,;\, {\bf s}_i^n) + I({\bf x}_i^n \, ; \, {\bf y}_i^n \, |\, {\bf s}_i^n)\big] + 
h({\bf y}_K^n) \nonumber \\
& & -\, h({\bf y}_K^n\, | \, {\bf x}_K^n) + (2K-1) n \epsilon_n\nonumber \\
& = & h({\bf y}_1^n)  - h({\bf y}_1^n\, | \, {\bf x}_1^n) + 
 \sum_{i=2}^{K-1} \big[h({\bf s}_i^n) - h({\bf s}_i^n \, | \, {\bf x}_i^n) + h({\bf y}_i^n \, | \, {\bf s}_i^n)
 - h({\bf y}_i^n \, | \, {\bf s}_i^n, {\bf x}_i^n) \big] + h({\bf y}_K^n) \nonumber \\
& & -\, h({\bf y}_K^n\, | \, {\bf x}_K^n) + (2K-1) n \epsilon_n. \label{sum_rate_eq_Ktx_step1}
\end{eqnarray}
Using the definition of the genie signals in \eqref{gen_sig_char_sum_rate_xc_ic_Ktx}, 
we note that the following are true 
\begin{eqnarray}
h({\bf y}_1^n\, | \, {\bf x}_1^n) & = & h({\bf s}_2^n) \nonumber \\
h({\bf s}_k^n \, | \, {\bf x}_k^n) & = & h({\bf s}_{k+1}^n), \qquad k = 2, \ldots, K-2.
\label{mi_s_k_x_k_eq_skp1}
\end{eqnarray}
Using \eqref{mi_s_k_x_k_eq_skp1} in \eqref{sum_rate_eq_Ktx_step1}, we have
\begin{eqnarray}
nS & \le & h({\bf y}_1^n) - h({\bf s}_{K-1}^n \, | \, {\bf x}_{K-1}^n) 
+ \sum_{i=2}^{K-1} \big[ h({\bf y}_i^n \, | \, {\bf s}_i^n) 
- h({\bf n}_i^n \, | \, {\bf s}_i^n, \, {\bf x}_i^n)  \big] \nonumber \\
& & + \, h({\bf x}_K^n + {\bf n}_K^n) - h({\bf n}_K^n) + (2K-1) n \epsilon_n   \nonumber \\
& \stackrel{(a)}{\le} & n  h({y}_{1G}) -{\bl h(h_K {\bf x}_K^n + {\bf n}_1^n)  }
+ \sum_{i=2}^{K-1} n \big[ h({y}_{iG} \, | \, {s}_{iG}) - h({n}_i)  \big] \nonumber \\
& & + \, h({\bf x}_K^n + {\bf n}_K^n)  - nh({n}_K) + (2K-1) n \epsilon_n  \nonumber \\
& \stackrel{(b)}{\le} & n h({y}_{1G}) 
+ \sum_{i=2}^{K-1} n \big[ h({y}_{iG} \, | \, {s}_{iG}) - h({n}_i)  \big]
+ \, n h({x}_{KG} + {n}_K)  \nonumber \\ 
& & - \,{\bl h(h_K \, {x}_{KG} + {n}_{1}) } - nh({n}_K) + (2K-1) n \epsilon_n, 
\label{sum_rate_eq_Ktx_step2} 
\end{eqnarray}
where $x_{iG} \sim \mathcal{N}(0, P_i)$, $y_{iG}$ denotes $y_i$ 
with $x_j = x_{jG}$, $\forall \,i, j$, $(a)$ follows from Lemma 1 in 
\cite{Veeravalli} and the fact that Gaussian inputs maximize the differential 
entropy for a given covariance constraint, $(b)$ follows from applying Lemma 1 
in \cite{Motahari} to the term 
$h({\bf x}_K^n + {\bf n}_K^n) -\, h(h_K \,  {\bf x}_K^n + {\bf n}_1^n)$, and using 
the condition $h_k^2 \; \le  \; 1$. 
Let $t_i$ denote the following quantity
\begin{eqnarray}
t_i & =  & 1  + \sum_{j \, = \, i}^K h_j^2 P_j.
\label{m21_t_i_defn}
\end{eqnarray}
Using \eqref{m21_t_i_defn}, we rewrite \eqref{sum_rate_eq_Ktx_step2} as
\begin{eqnarray}
nS & \le & \frac{n}{2} \log \pi e (t_2 + P_1) + \frac{n}{2} \sum_{i \, = \, 2}^{K-1} \log \left[\frac{(1 + P_i) t_i - h_i^2 P_i^2}{t_i} \right]
+ \frac{n}{2}\log \pi e(1 + P_K)  \nonumber \\
& & - \, \frac{n}{2}\log \pi e (t_K) - \frac{n}{2}  \log \pi e + (2K-1) n \epsilon_n  \nonumber \\
& = & 0.5\log \left(1 + \frac{P_1}{t_2}\right) 
+ 0.5n \sum_{i \, = \, 2}^{K-1} \log \left[\frac{(1 + P_i) \, t_i - h_i^2 P_i^2}{t_{i+1}} \right] \nonumber \\
& & +\, 0.5n \log (1 + P_K)  + (2K-1) n \epsilon_n.
\label{xc_as_ic_char_sum_rate_gen_aid_Ktx}
\end{eqnarray}

The achievable sum-rate of a scheme that employs Gaussian codebooks and treats interference as noise is 
given by 
\begin{eqnarray}
S_{ach} & {=}  & 0.5\log \left(1 + \frac{P_1}{1 + \sum_{j=2}^K h_j^2 P_j} \right ) + 0.5 \sum_{i=2}^K \log(1 + P_i)
\nonumber \\
& = & 0.5 \log \left(1 + \frac{P_1}{t_2}\right) + 0.5 \sum_{i=2}^K \log(1 + P_i).
\label{ach_sum_rate_tin}
\end{eqnarray} 
Subtracting \eqref{ach_sum_rate_tin} from \eqref{xc_as_ic_char_sum_rate_gen_aid_Ktx}, the gap $\delta$ between the 
genie-aided outer bound and the achievable sum-rate is given by
\begin{eqnarray}
\delta & = & \sum_{i \, = \, 2}^{K-1}0.5  \log \left[\frac{(1 + P_i) \, t_i - h_i^2 P_i^2}{t_{i+1} (1 + P_i)} \right] 
+ (2K-1) \epsilon_n  \nonumber \\
& =& \sum_{i \, = \, 2}^{K-1} 0.5 \log \left[\frac{(1 + P_i) \, (h_i^2 P_i + t_{i+1}) - h_i^2 P_i^2}{t_{i+1} (1 + P_i)} \right] 
+ (2K-1) \epsilon_n  \nonumber \\
& = & \sum_{i \, = \, 2}^{K-1} 0.5  \log \left[ 1 + \frac{h_i^2 P_i}{t_{i+1} (1 + P_i)} \right]
+ (2K-1) \epsilon_n  \label{delta_bnd_m21_xc_1} \\
& \stackrel{(c)}{\le} & K/2 - 1 + (2K-1) \epsilon_n,
\label{delta_bnd_m21_xc_2}
\end{eqnarray} 
where we have used $h_i^2 P_i \le (1 + P_i)$ and $t_{i+1} \ge 1$ to write $(c)$. As $n \rightarrow \infty$, $\epsilon_n \rightarrow 0$ and therefore {\bl $\delta \le K/2 - 1$}. 
We note that if $K = 3$,  $\delta \le 0.5$, 
which implies that the total gap is within half a bit.
\end{IEEEproof}

{\bl
\begin{remark}
A similar result is proved for the $K \times K$ XC in \cite{Geng_XC}, where they show that
under certain channel conditions, strategy $\mathcal{M}1$, i.e., treating interference as noise at the receivers is sum generalized degrees-of-freedom (GDoF) optimal and also
achieves a constant gap to the sum-rate capacity. This result can be specialized to the many-to-one XC, and after some manipulations, the channel conditions in \cite[Theorem 2]{Geng_XC} essentially boil down to sub-region \eqref{sub_reg_oper_xc_ic_Ktx}, where it is shown that the  gap from the sum-rate capacity is within $\frac{K}{2} \log_2 \big[K (K + 1)\big]$ bits. Note that the gap from 
the sum-rate capacity is larger than that in Theorem \ref{thm_m21_xc_char_1_bit_per_user}, owing to the fact that the bounding techniques as well as
the results in \cite{Geng_XC} are applicable to the general fully connected $K \times K$ XC. 
\end{remark}
}

\section{{\bl$K$-user Gaussian} many-to-one Interference channel}
\label{sec_gauss_m21_ic}

In this section, we observe some implications of the above results for the {\bl $K$-user} Gaussian many-to-one IC.
The system model for  {\bl the $K$-user} Gaussian many-to-one IC written in standard form is 
same as that of the many-to-one XC shown in Fig. \ref{many_2_1_XC_Ktx_std_form}, with the exception that the 
cross messages are now absent, i.e., $W_{1j} = \phi$, 
$j = 2, \ldots, K$.  From Fano's inequality, we have
\begin{eqnarray}
H(W_{ii} \, | \,{\bf y}_i^n) & \le & n \epsilon_n, \qquad   \label{m21_ic_fano_ineq}
\end{eqnarray}

Note that in the Gaussian many-to-one IC, all transmitters excluding 
the first cause interference for the reception of the intended signal at receiver 1.
Transmission strategies can similarly be defined for the Gaussian many-to-one 
IC and lead to characterization of sum-rate capacity in some sub-regions.
The strategies naturally involve a combination of decoding a part of the 
interference and treating the rest of the interference as noise. 
This leads to the following definition.

\begin{definition}
In Strategy $\mathcal{MI}{\small k}$, 
interference resulting from transmissions from $k-1$ transmitters is 
decoded and canceled at receiver 1, while the rest of the interference from 
other transmitters is treated as noise, $k \in \{1, \ldots, K\}$.
\end{definition}

{\bl Thus, strategy $\mathcal{MI}1$ refers to the case where interference 
from all transmitters is treated as noise at receiver 1.} Strategy $\small \mathcal{MI}K$ refers to the case where interference 
from all transmitters is decoded and canceled at receiver 1.

\subsection{Conditions for the sum-rate optimality of strategy $\mathcal{MI}k$}
We use sum-rate as the criterion of optimality for 
evaluating the strategies. In the $K \times K$ Gaussian many-to-one 
XC studied in Section \ref{subsec_cond_sum_rate_opt_m21_xc}, we characterized 
the sum-rate optimality 
of strategies $\mathcal{M}1$, $\mathcal{M}2$ and also characterized 
the gap from the optimality of strategy $\mathcal{M}3$. However, 
in the Gaussian many-to-one IC, we characterize the sum-rate 
optimality of all strategies, $\mathcal{MI}1$ to $\mathcal{MI}K$.
Without loss of generality, we assume that strategy  $\mathcal{MI}k$ 
refers to decoding interference from transmitters 2 through $k$, while 
interference from transmitters $k+1$ through $K$ is treated as noise.
The result for the general case where interference from any subset of transmitters 
of cardinality $k-1$ is decoded can be obtained from a reordering of the transmitters
without any loss in sum-rate.

Let $\mathcal{Q}$ denote the set of integers $\{2, 3, \ldots, k\}$.
Let $\pi^{\mathcal{Q}}$ denote any permutation of the set $\mathcal{Q}$ 
with $\pi^{\mathcal{Q}}(i)$ denoting the $i$th element of the permutation.
We have the following result on the sum-rate 
optimality of strategy $\mathcal{MI}k$, $k \in  \{1, \ldots, K\}$.

\begin{theorem}
For a {\bl $K$-user Gaussian}  many-to-one IC satisfying the following channel conditions
\begin{eqnarray}
h_{\pi^{\mathcal{Q}}(i)}^2 & \ge & 1 + P_1 + \sum_{\substack{j \in \pi^{\mathcal{Q}} \\ j > i}} 
h_{\pi^{\mathcal{Q}}(j)}^2 P_{\pi^{\mathcal{Q}}(j)} + \sum_{j = k+1}^K 
h_j^2 P_j, \qquad i = 1, \ldots, k -1, \label{cond_m21_ic_opt_strat_Mk_1}  \\
\sum_{j=k+1}^{K} h_j^2 & \le &  1, 
\label{cond_m21_ic_opt_strat_Mk_2}
\end{eqnarray}
for some permutation $\pi^{\mathcal{Q}}$, decoding interference from 
transmitters 2 to $k$ and treating interference from the rest of 
the transmitters as noise achieves the sum-rate capacity, and is given by 
\begin{eqnarray}
S & \le & \log \bigg(1 + \frac{P_1}{1 + \sum_{j=k+1}^{K} h_j^2 \, P_j} \bigg) + 
\sum_{i=2}^{K}  \log( 1 + P_i). \label{m21_ic_sum_rate_cap_strat_mi_k} \nonumber 
\end{eqnarray}
\label{thm_m21_ic_sum_cap_strat_Mk}
\end{theorem}

\begin{IEEEproof}
First, we prove the converse.
Let a genie provide the following genie signals to receiver 1
\begin{eqnarray}
{\bf s}_1 & = &   (x_2, x_3, x_4, \ldots, x_{k}). \nonumber 
\end{eqnarray}

The sum-rate of the genie-aided channel is given by
\begin{eqnarray}
nS & = &  \sum_{i\,=\,1}^K H(W_{ii})  \nonumber \\
& = & {\bl I(W_{11} \, ; \, {\bf y}_1^n,  {\bf s}_1^n) +
 \sum_{i=2}^{K} I(W_{ii} \, ; \, {\bf y}_i^n) + H(W_{11} \,|\, {\bf y}_1^n, \, {\bf s}_1^n) + \sum_{i=2}^{K} H(W_{ii} \,|\, {\bf y}_i^n)} \nonumber \\
& \stackrel{(a)}{\le} &  {\bl I({\bf x}_1^n \,;\, {\bf y}_1^n, \, {\bf s}_1^n) 
+ \sum_{i=2}^{K} I({\bf x}_i^n \, ; \, {\bf y}_i^n)
+ \sum_{i=1}^{K} H(W_{ii} \,|\, {\bf y}_i^n)} \nonumber \\
& \stackrel{(b)}{\le} &  I({\bf x}_1^n \,;\, {\bf s}_1^n) + I({\bf x}_1^n \,;\, {\bf y}_1^n \, | \, {\bf s}_1^n) 
+ \sum_{i=2}^{K} I({\bf x}_i^n \, ; \, {\bf y}_i^n) +
nK \epsilon_n  \nonumber \\
& \stackrel{(c)}{\le} & I({\bf x}_1^n \,;\, {\bf y}_1^n \, | \, {\bf s}_1^n) 
+ \sum_{i=2}^{K} I({\bf x}_i^n \, ; \, {\bf y}_i^n)
+ nK \epsilon_n  \nonumber \\
& = & h({\bf y}_1^n \, | \,  {\bf s}_1^n) - h({\bf y}_1^n \, | \,  {\bf s}_1^n, {\bf x}_1^n) 
+ \sum_{i=2}^{K} \left[h({\bf y}_i^n) - h({\bf y}_i^n  \, | \,{\bf x}_i^n)  \right] + n K \epsilon_n \nonumber \\
& = & h\bigg({\bf x}_1^n + \sum_{j=k+1}^{K} h_j \, {\bf x}_j^n + {\bf n}_1^n\bigg) 
- h\bigg(\sum_{j=k+1}^{K} h_j \, {\bf x}_j^n + {\bf n}_1^n\bigg) + \sum_{i=2}^{K} \left[h({\bf y}_i^n) 
- h({\bf n}_i^n) \right] \nonumber + nK \epsilon_n \\
& = & h({\bf y}_1^n \, | \,  {\bf s}_1^n) - h({\bf y}_1^n \, | \,  {\bf s}_1^n, {\bf x}_1^n) 
+ \sum_{i=2}^{K} \left[h({\bf y}_i^n) - h({\bf y}_i^n  \, | \,{\bf x}_i^n)  \right] + n K \epsilon_n \nonumber \\
& = & {\bl h\bigg({\bf x}_1^n + \sum_{j=k+1}^{K} h_j \, {\bf x}_j^n + {\bf n}_1^n\bigg) 
- h\bigg(\sum_{j=k+1}^{K} h_j \, {\bf x}_j^n + {\bf n}_1^n\bigg) + \sum_{i=2}^{k} h({\bf y}_i^n)  } \nonumber \\ 
& &  {\bl  + \sum_{i=k+1}^{K} h({\bf y}_i^n)  
-  \sum_{i=2}^{K} h({\bf n}_i^n) \nonumber + nK \epsilon_n } \\
 & \stackrel{(d)}{\le} & n h\bigg({x}_{1G} + \sum_{j=k+1}^{K} h_j \, {x}_{jG} + {n}_1\bigg)  
- h\bigg(\sum_{j=k+1}^{K} h_j \, {\bf x}_j^n + {\bf n}_1^n \bigg) + \sum_{i=2}^{k} n h({y}_{iG})
\nonumber \\
& & + \,\sum_{i=k+1}^{K} h({\bf x}_i^n + {\bf n}_i^n) -  \sum_{i=2}^{K} n h({n}_i) + nK \epsilon_n\nonumber \\
& \stackrel{(e)}{\le} & n h\bigg({x}_{1G} + \sum_{j=k+1}^{K} h_j \, {x}_{jG} + {n}_1\bigg)  
+ \sum_{i=2}^{K} n h({y}_{iG}) - n h\bigg(\sum_{j=k+1}^{K} h_j \, {x}_{jG} + {n}_1\bigg) \nonumber \\
& & - \, \sum_{i=2}^{K} n h({n}_i) + nK \epsilon_n \nonumber \\
& = & n I({x}_{1G} \,;\, {y}_{1G}, \, {s}_{1G}) + 
\sum_{i=2}^{K} n I({x}_{iG} \, ; \, {y}_{iG}) + nK \epsilon_n  \nonumber \\
& = & \frac{n}{2} \log \bigg(1 + \frac{P_1}{1 + \sum_{j=k+1}^{K} h_j^2 \, P_j} \bigg) 
+ \sum_{i=2}^{K} \frac{n}{2} \log( 1 + P_i) + nK \epsilon_n, \label{m21_ic_sum_rate_bound_strat_mi_k}
\end{eqnarray}
where $(a)$ follows from the fact that removing conditioning cannot reduce the 
conditional entropy, $(b)$ follows from \eqref{m21_ic_fano_ineq},  $(c)$ follows from 
the independence of ${\bf s}_1^n$ and ${\bf x}_1^n$, 
$(d)$ follows since Gaussian inputs maximize differential entropy 
for given covariance constraints, and $(e)$ follows from the application of 
Lemma \ref{lemma_li} to bound the term $\sum_{i=k+1}^{K} h({\bf x}_i^n + {\bf n}_i^n) - 
h\bigg(\sum_{j=k+1}^{K} h_j \, {\bf x}_j^n + {\bf n}_1^n \bigg)$, under 
the condition $\sum_{j=k+1}^{K} h_j^2 \le 1$.

For  achievability, note that the sum-rate outer bound in 
\eqref{m21_ic_sum_rate_bound_strat_mi_k} can be achieved by using 
Gaussian inputs, decoding and canceling interference from transmitters 2 to $k$ and 
treating interference from transmitters $k+1$ to $K$ as noise. 
Assume Gaussian inputs are used at each transmitter, i.e., $x_i = x_{iG}$, 
$i = 1, \ldots, K$.
The order in which the signals from transmitters 2 to $k$ 
are decoded at receiver 1 determines the channel conditions 
that must be satisfied for achievability.  
Here, we use $\pi^{\mathcal{Q}}$ to denote the decoding order at receiver 1, 
with $\pi^{\mathcal{Q}}(i)$ decoded and canceled out before decoding $\pi^{\mathcal{Q}}(j)$ for $i < j$.

For ease of presentation, we use  $\pi^{\mathcal{Q}} = \{2, 3, \ldots, k\}$  
with no permutation, i.e., $x_{2G}$ is decoded and cancelled out before decoding 
$x_{3G}$ and so on.

Notice that, 
\begin{eqnarray}
I(x_{2G} \, ; \, y_{1G}) = I\left(x_{2G} \, ; \, x_{2G} + \frac{x_1 + \sum_{j=3}^{K} h_j \,x_{jG} + n_1}{h_2}\right) 
& \ge & I(x_{2G} \, ; \, y_{2G}), \nonumber 
\end{eqnarray}
if $h_2^2 \ge  1 + P_1 + \sum_{j=3}^{K} h_j^2 P_j$.
Similarly, for some $2 < l \le k$, we have
\begin{eqnarray}
I(x_{lG} \, ; \, y_{1G}\, |\, x_{2G}, \ldots, x_{(l-1)G}) = I\left(x_{lG} \, ; \, x_{lG} + \frac{x_1 + \sum_{j=l+1}^{K} h_j \,x_{jG} + n_1}{h_l}\right) 
& \ge & I(x_{lG} \, ; \, y_{lG}), \nonumber
\end{eqnarray}
if $h_l ^2 \ge  1 + P_1 + \sum_{j=l+1}^{K} h_j^2 P_j$.
Combining the above channel conditions, we have 
\begin{eqnarray}
h_i^2 & \ge &   1 + P_1 + \sum_{j=i+1}^{K} h_j^2 P_j, \qquad i = 2, \ldots, k.
\end{eqnarray}

Thus, \eqref{cond_m21_ic_opt_strat_Mk_1} represents the above condition for a 
random permutation of $\mathcal{Q}$ and \eqref{cond_m21_ic_opt_strat_Mk_2} 
is needed to prove the sum-rate outer bound in \eqref{m21_ic_sum_rate_bound_strat_mi_k}.
This completes the proof of the theorem.
\end{IEEEproof}

\subsection{Conditions for sum-rate of strategy of $\mathcal{MI}1$
to be within $K/2-1$ bits from sum-rate capacity}

Here, we obtain a region for the Gaussian many-to-one IC, where the 
sum-rate capacity can be characterized to within $K/2-1$ bits .
In Theorem \ref{thm_oper_m21_xc_as_ic_Ktx}, we showed that in 
sub-region \eqref{sub_reg_oper_xc_ic_Ktx}, the Gaussian many-to-one XC can be operated 
as a Gaussian many-to-one IC without loss of sum-rate. Further, in Theorem \ref{thm_m21_xc_char_1_bit_per_user}, 
we showed that in the above sub-region, the sum-rate of strategy $\mathcal{M}1$ 
is within $K/2-1$ bits from the sum-rate capacity. 
Notice that strategy $\mathcal{M}1$ for the Gaussian many-to-one XC, which 
involves using Gaussian codebooks and treating interference as noise, corresponds 
to strategy $\mathcal{MI}1$ in many-to-one IC. Since the sum-rate capacity of 
the Gaussian many-to-one XC forms an outer bound on the sum-rate capacity of 
Gaussian many-to-one IC, we conclude that strategy $\mathcal{MI}1$ is within 
$K/2-1$ bits from the sum-rate capacity of Gaussian many-to-one IC in sub-region 
\eqref{sub_reg_oper_xc_ic_Ktx}.

In the following theorem, we show that strategy $\mathcal{MI}1$ achieves 
a rate point that is within $K/2-1$ bits from the sum-rate capacity of 
Gaussian many-to-one IC in a region that is much larger than 
sub-region \eqref{sub_reg_oper_xc_ic_Ktx}.
Let $\mathcal{S}$ denote the set of integers $\mathcal{S} = \{2, 3, \ldots, K \}$. 
Let $\pi^S$ denote any permutation of the elements of the set $\mathcal{S}$, 
with $\pi^S(k)$ denoting the $k$th element of the permutation.

\begin{theorem}
{\bl For the $K$-user Gaussian many-to-one IC},
the rate point achieved by using Gaussian codebooks and treating interference 
as noise is within $K/2-1$ bits from the sum-rate capacity of Gaussian 
many-to-one IC in the following sub-regions
\begin{eqnarray}
h_{\pi^S(i)}^2 & \le &  \bigg(1 + \frac{1}{P_{\!\pi^S(i)}}\bigg) {\bl \bigg( 1 \; \,+ } \sum_{j \, = \, \pi^S(i+1)}^{{\bl \pi^S(K-1)}} h_j^2 P_j \bigg), 
\quad  \qquad i = 1, \ldots, K-2, \nonumber \\
h_{\pi^S(K-1)}^2  & \le & 1.
\label{sub_reg_sum_char_m21_ic}
\end{eqnarray}
\label{thm_m21_ic_char_1_bit_per_user_Ktx}
\end{theorem}

\begin{IEEEproof}
Without loss of generality, we assume $\pi^S = \mathcal{S}$, i.e., no permutation of the elements 
of the set $\mathcal{S}$ is assumed. Thus, $\pi^S(1) = 2$, $\pi^S(2) = 3$ and so on till 
$\pi^S(K-1) = K$.

Let a genie provide the side-information given in \eqref{gen_sig_char_sum_rate_xc_ic_Ktx} 
to receiver $i$, $i = 2, \ldots, K-1$. The sum-rate of the genie-aided channel is bounded as 
\begin{eqnarray}
nS & = &  \sum_{i\,=\,1}^K H(W_{ii}) \nonumber \\
& = &  I(W_{11} \,;\, {\bf y}_1^n) 
+ \sum_{i=2}^{K-1} I(W_{ii} \, ; \, {\bf y}_i^n, \,  {\bf s}_i^n)
+ I(W_{KK} \,;\, {\bf y}_K^n) + {\bl H(W_{11} \,|\, {\bf y}_1^n)} \nonumber \\
& & {\bl +\, \sum_{i=2}^{K-1} H(W_{ii} \,|\, {\bf y}_i^n, \, {\bf s}_i^n) 
+  H(W_{KK} \,|\, {\bf y}_K^n) } \nonumber \\
& \stackrel{(a)}{\le} &  I({\bf x}_1^n \,;\, {\bf y}_1^n) 
+ \sum_{i=2}^{K-1} I({\bf x}_i^n \, ; \, {\bf y}_i^n, \,  {\bf s}_i^n)
+ I({\bf x}_K^n \,;\, {\bf y}_K^n) 
{\bl +\, \sum_{i=1}^{K} H(W_{ii} \,|\, {\bf y}_i^n) 
} \nonumber \\
& \stackrel{(b)}{\le} & I({\bf x}_1^n \,;\, {\bf y}_1^n) 
+ \sum_{i=2}^{K-1} I({\bf x}_i^n \, ; \, {\bf y}_i^n, \,  {\bf s}_i^n)
+ I({\bf x}_K^n \,;\, {\bf y}_K^n) + nK \epsilon_n, \label{sum_rate_m21_IC_gen_aid_Ktx} 
\end{eqnarray}
where $(a)$ follows from the fact that removing conditioning cannot reduce the 
conditional entropy, and $(b)$ follows from \eqref{m21_ic_fano_ineq}. 
We recognize that \eqref{sum_rate_m21_IC_gen_aid_Ktx} is 
similar to \eqref{sum_rate_eq_m21_IC_gen_aid_Ktx}. Notice that the constraint $h_i^2 \le 1$, 
needed to write \eqref{sum_rate_eq_m21_IC_gen_aid_Ktx} for the many-to-one XC 
is not required in the case of many-to-one IC.

By following essentially the same set of steps as in the 
Theorem \ref{thm_m21_xc_char_1_bit_per_user}, and letting $\delta'$ 
denote the gap between the genie-aided outer bound and the achievable sum-rate for the many-to-one IC, 
it follows that $\delta'$ is bounded by \eqref{delta_bnd_m21_xc_1} if $h_k^2 \le 1$.
Note that the condition $h_k^2 \le 1$ is required to write the inequality \eqref{sum_rate_eq_Ktx_step2} in  Theorem 
 \ref{thm_m21_xc_char_1_bit_per_user}.

Using \eqref{delta_bnd_m21_xc_1}, we conclude that for a gap of {\bl $K/2-1$ bits}, 
if $h_i^2 P_i \le t_{i+1} (1 + P_i)$, along with $h_k^2 \le 1$, then 
{\bl $\delta' \le (K/2-1) + K \epsilon_n \Rightarrow \delta' \le K/2 - 1$}.
We  again note that for $K=3$, $\delta' \le 0.5$, {\bl implying that a total gap 
of within half a bit is }obtained from the sum-rate capacity.
The above conditions can be rewritten as 
\begin{eqnarray}
h_{i}^2 & \le &  \bigg(1 + \frac{1}{P_{i}}\bigg) {\bl \bigg( 1 \;+ } \sum_{j \, = \, i+1}^{K} h_j^2 P_j \bigg), \qquad  \quad i = 2, \ldots, K -1, \nonumber \\
h_{K}^2  & \le & 1.  \nonumber 
\end{eqnarray}
Note that the above region is much larger than sub-region \eqref{sub_reg_oper_xc_ic_Ktx}, i.e., $h_i^2  \le 1,  i = 2, \ldots, K$, obtained for the many-to-one XC in Theorem \ref{thm_m21_xc_char_1_bit_per_user}.
We illustrate the above region for $K=3$ 
in Fig. \ref{plot_a_b_graph_SNR_0dB_m21_IC}.

The general case for any permutation $\pi^S$ of $\mathcal{S}$ can be proved by giving the following 
genie signal to receiver $\pi^S(i)$, $ i = 1, \ldots, K-2$
\begin{eqnarray}
s_{\pi^S(i)} & = & \sum_{j\,=\,\pi^S(i)}^K h_j \, x_j + n_1.
\nonumber 
\end{eqnarray}
and following the steps given above.
\end{IEEEproof}

\begin{remark}
In \cite{Bresler}, inner and outer bounds to the  capacity region of the 
Gaussian many-to-one IC are presented. The inner bound is based on an 
achievable scheme which uses lattice codes for alignment of interfering signals 
at receiver 1. The outer bound is proved by giving an appropriately chosen side 
information to receiver 1. It is shown that the gap between the inner and outer 
bounds is approximately $5K \log K$ bits per user with $K + 1$ users in the system. 
In Theorem \ref{thm_m21_ic_char_1_bit_per_user_Ktx}, we have strengthened the 
above result for the sub-region in \eqref{sub_reg_sum_char_m21_ic}, 
by showing that using Gaussian codebooks and treating interference as noise is 
within $K/2-1$ bits from the sum-rate capacity of the many-to-one IC.
\end{remark}

\section{Numerical results}
\label{sec_num_res}

In this section, we illustrate the regions where the derived channel conditions
are satisfied for each strategy. For ease of presentation, we consider the 
$3 \times 3$ many-to-one XC for evaluating the strategies.

First, we numerically analyze the sum-rate outer
bound for the optimality of strategy $\small \mathcal{M}3$, given in Theorem
\ref{thm_MAC_rx1}. 
Let the gap between the sum-rate outer bound and the achievable sum-rate of 
strategy $\small \mathcal{M}3$ given in \eqref{sum_rate_gap} be denoted by 
$\Delta$. Using \eqref{sum_rate_gap} and solving for $\rho$ in terms of $\Delta$, we get
\begin{eqnarray}
\rho^2 & \le & \frac{2^{2\Delta} - 1}{2^{2\Delta} - 1/(1 + b^2 \,P_3)}.
\label{one_2_many_rho_eqn}
\end{eqnarray}

\begin{figure}[t]
\centering
\includegraphics[width=0.65\columnwidth]{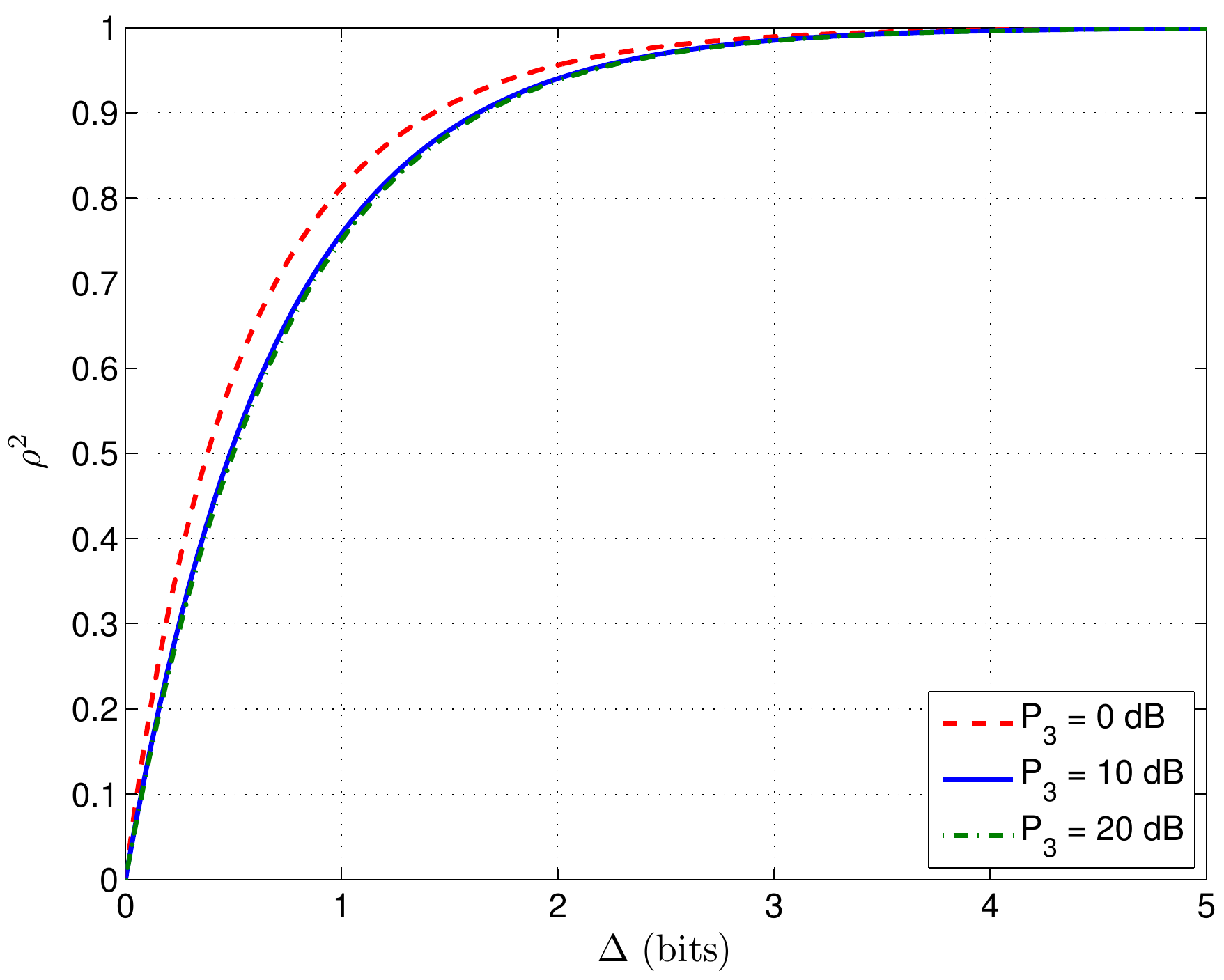}
\caption{Variation of $\rho^2$ as a function of the gap $\Delta$ in bits. $b
= 1.5$.}
\label{plot_rho_vs_delta}
\end{figure}

In Fig. \ref{plot_rho_vs_delta}, we plot $\rho^2$ as a function of $\Delta$ for different 
values of $P_3$ for fixed value of $b = 1.5$. It can be observed that $\rho^2$ is a monotonically increasing function 
of $\Delta$. Thus, to obtain a lower gap from the outer bound, a lower value of $\rho^2$  
must be chosen. This in turn makes the sub-region in \eqref{many_2_one_strt2_sub_reg} 
smaller. This relationship is explored further is the next two plots.

\begin{figure}[t]
\centering
\includegraphics[width=0.8\columnwidth]{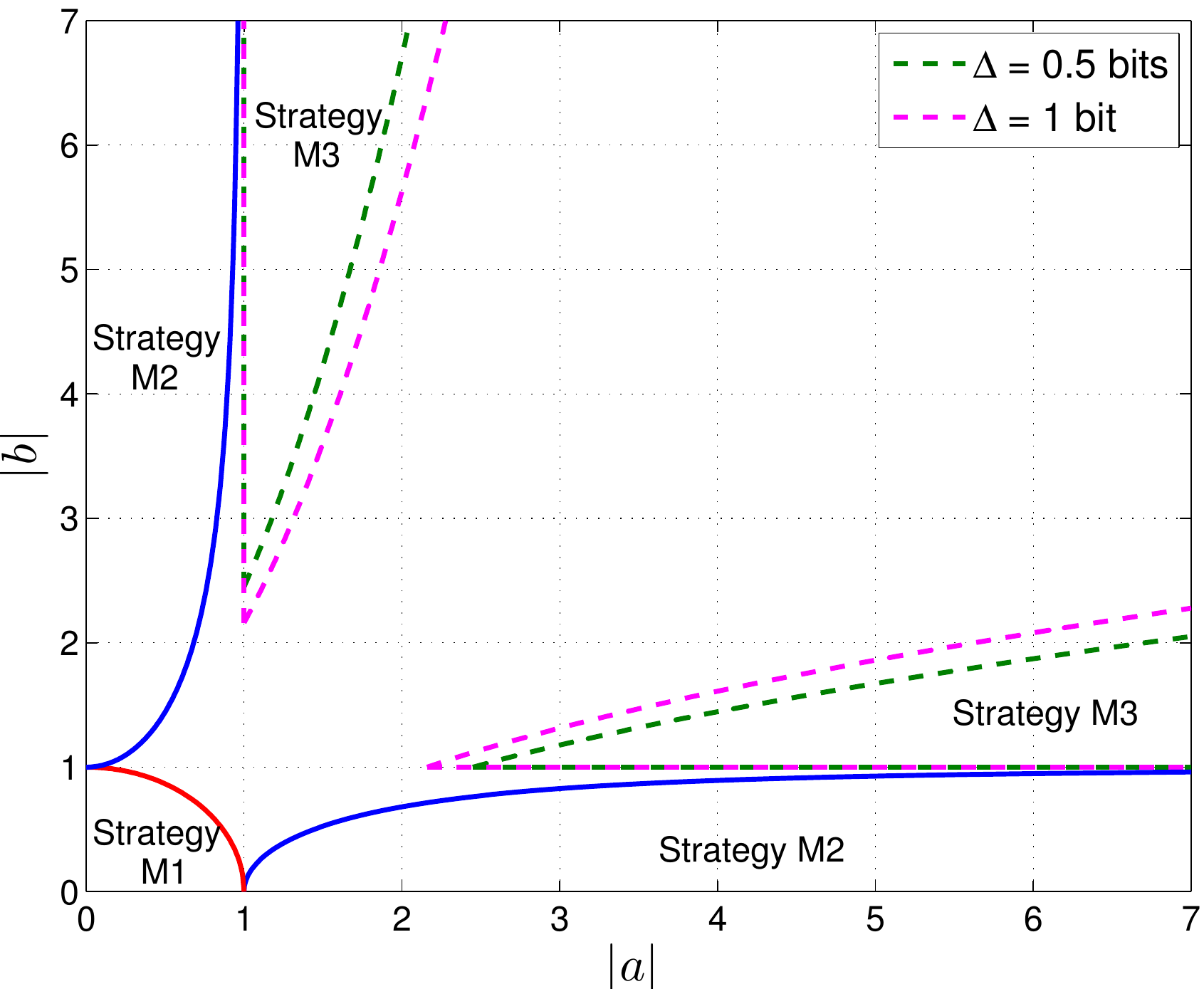}
\caption{A plot of the channel conditions in Table
\ref{table_many21_sum_results} for a $3 \times 3$ many-to-one XC 
for the three strategies.  $P_1 = P_2 = P_3 = 0$ dB.}
\label{plot_a_b_graph_many_2_1}
\end{figure}

\begin{figure}[t]
\centering
\includegraphics[width=0.8\columnwidth]{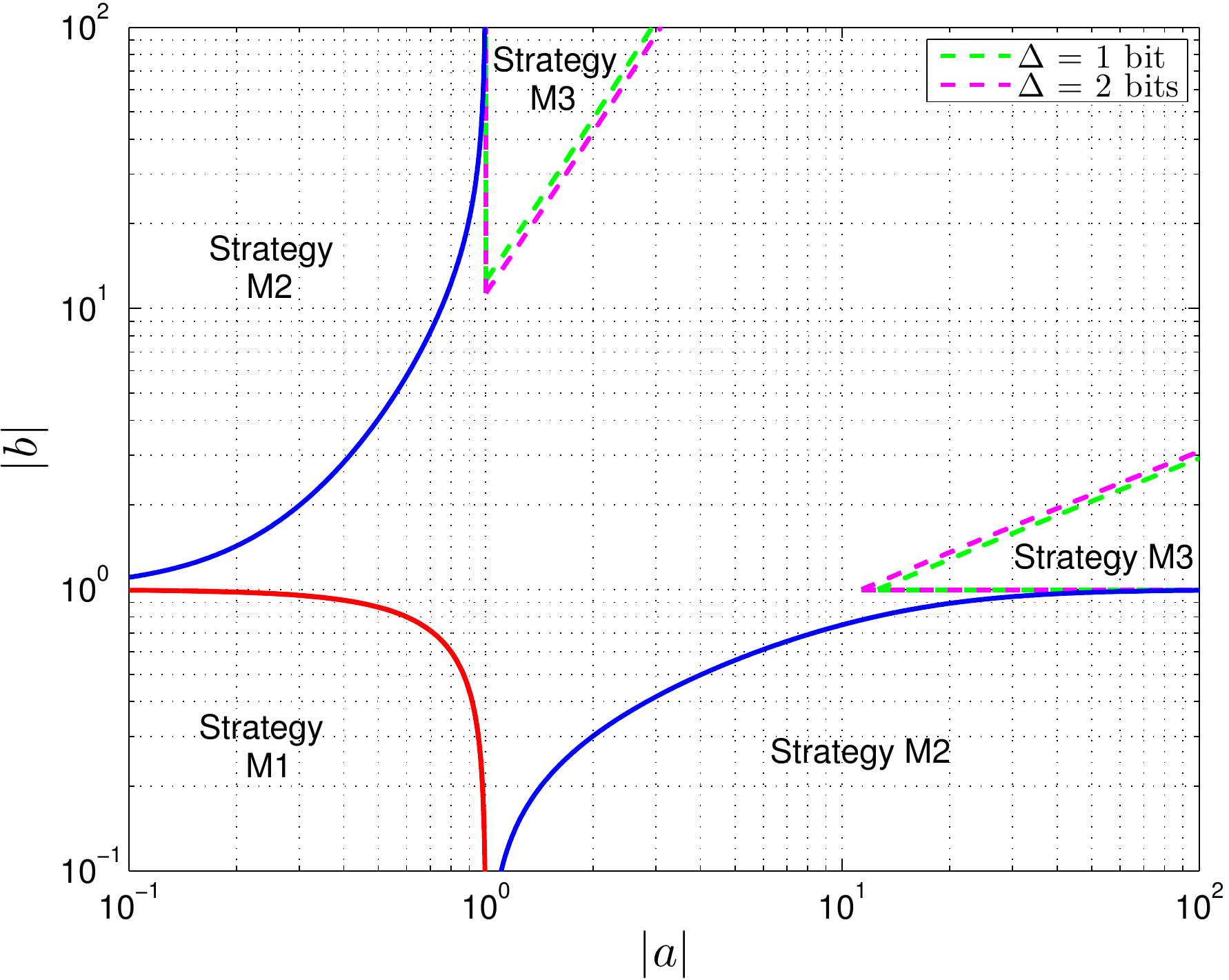}
\caption{A plot of the channel conditions in Table
\ref{table_many21_sum_results} for a $3 \times 3$ many-to-one XC 
for the three strategies.  $P_1 = P_2 = P_3 = 10$ dB.}
\label{plot_a_b_graph_SNR_10dB}
\end{figure}

In Fig. \ref{plot_a_b_graph_many_2_1} and Fig. \ref{plot_a_b_graph_SNR_10dB}, we plot the sub region in
\eqref{many_2_one_strt2_sub_reg} for the sum-rate optimality of strategy
 $\small \mathcal{M}3$ as a graph in the $|a|\!-\!|b|$ 
plane for various values of $\Delta$, along with the sub-regions in Table
\ref{table_many21_sum_results} for strategies
$\small \mathcal{M}1$ and $\small \mathcal{M}2$. We assume $P_1 = P_2 = P_3 = 0$ dB.
As mentioned above, the sub-region in \eqref{many_2_one_strt2_sub_reg} shrinks 
for increasing values of $\Delta$.

\begin{table}[t]
\centering
{
\renewcommand{\arraystretch}{2.5}
\renewcommand{\tabcolsep}{.4cm}
\begin{tabular}{|>{\normalsize}c|>{\normalsize}c|}
\hline
 Strategy &  {Channel conditions} \\
\hline
\normalsize {$\normalsize \mathcal{MI}1$ } & 
$a^2 + b^2 \le 1$ \\
\hline
\normalsize {$\normalsize \mathcal{MI}2$ } & (i)
$a^2 \ge 1 + P_1 + b^2 P_3, \;\,b^2 \le 1$ \normalsize   
\\
\cline{2-2}
&  (ii) 
$b^2 \ge 1 + P_1 + a^2 P_2, \;\,a^2 \le 1$   \\
\hline
\normalsize {$\normalsize \mathcal{MI}3$ } &  (i) 
$a^2 \ge 1 + P_1 + b^2 P_3, \;\,b^2 \ge 1 + P_1$ 
\\
\cline{2-2}
&(ii)  $b^2 \ge 1 + P_1 + a^2 P_2, \;\,a^2 \ge 1 + P_1$ 
 \\
\hline
\end{tabular}
}
\caption{Sum-rate capacity results for a $3 \times 3$ many-to-one IC in 
Theorem \ref{thm_m21_ic_sum_cap_strat_Mk}.}
\label{table_many21_IC_results}
\end{table}

\begin{figure}[t]
\centering
\includegraphics[width=0.7\columnwidth]{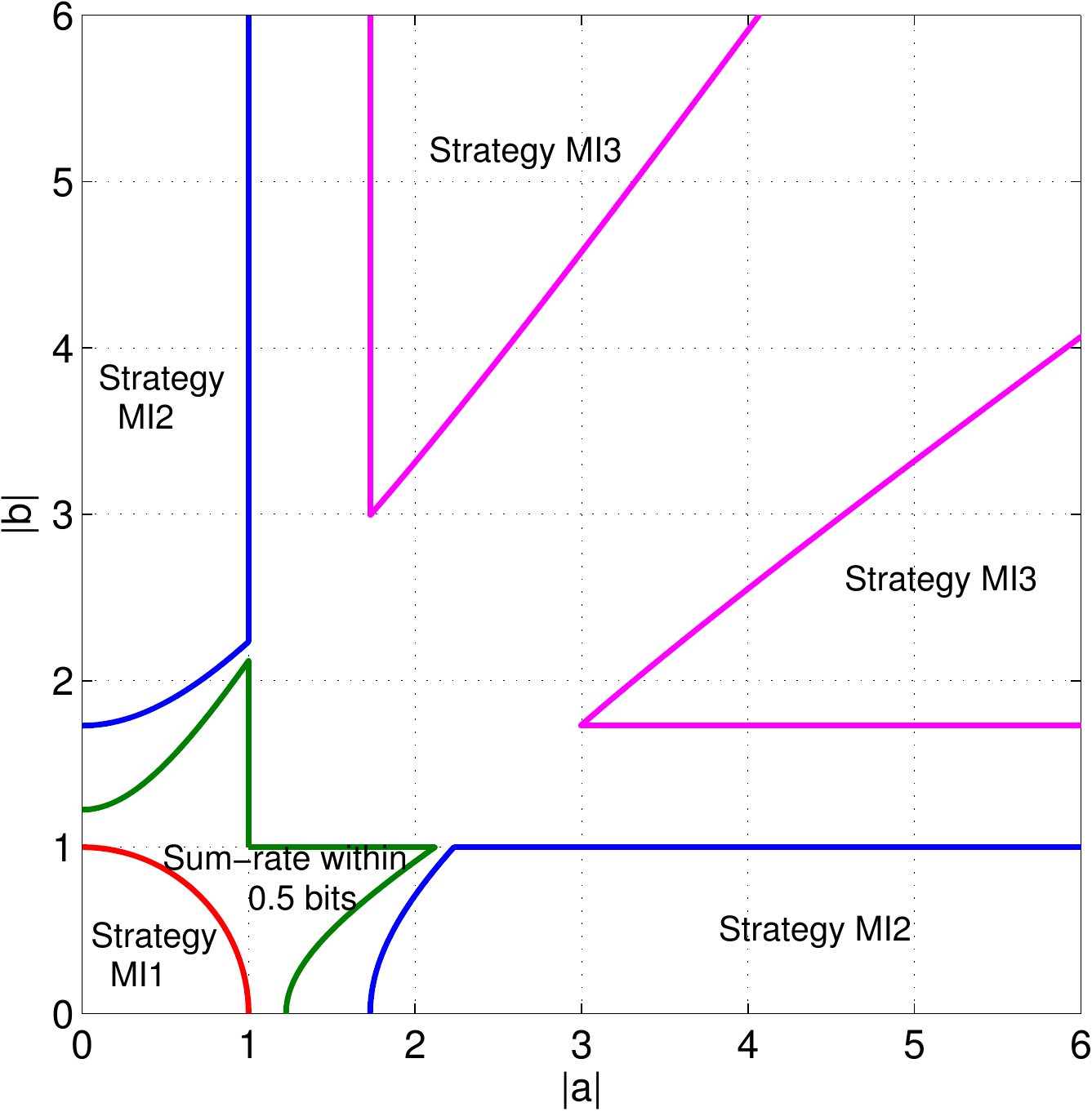}
\caption{A plot of the channel conditions in Theorem \ref{thm_m21_ic_sum_cap_strat_Mk} 
(summarized in Table \ref{table_many21_IC_results}) and 
Theorem \ref{thm_m21_ic_char_1_bit_per_user_Ktx} for a $3 \times 3$ many-to-one IC. 
$P_1 = P_2 = P_3 = 3$ dB.}
\label{plot_a_b_graph_SNR_0dB_m21_IC}
\end{figure}

{\bl In Fig. \ref{plot_a_b_graph_SNR_0dB_m21_IC}, 
we plot the characterization of sum-rate capacity for the Gaussian many-to-one 
IC obtained in Theorem \ref{thm_m21_ic_char_1_bit_per_user_Ktx} for a  
$3 \times 3$ many-to-one IC.} Also plotted are the 
channel conditions determined in Theorem \ref{thm_m21_ic_sum_cap_strat_Mk} for strategies $\mathcal{MI}1$, 
$\mathcal{MI}2$, and $\mathcal{MI}3$ to achieve sum-rate capacity.
For $K=3$, and using same notation as in many-to-one XC with $a = h_2$, 
$b = h_3$, sub-region \eqref{sub_reg_sum_char_m21_ic} becomes 
\begin{enumerate}
\item[(i)] \indent \indent \indent $a^2  \; \le \; (1 + b^2 P_3)\left(\displaystyle 1 + \frac{1}{P_2}\right)      ; \quad b^2 \le 1$
\item[(ii)] $b^2  \; \le \; (1 + a^2 P_2)\left(\displaystyle 1 + \frac{1}{P_3}\right)   ; \quad a^2 \le 1$.
\end{enumerate}
The above region is illustrated in the figure for $P_1 = P_2 = P_3 = 3$ dB. 
As mentioned earlier, for $K=3$, the total gap between 
the sum-rate of strategy $\mathcal{MI}1$ and the sum-rate capacity 
of the $3 \times 3$ many-to-one IC is less than one bit.
Thus, as long as the channel coefficients lie within this region, 
the sum-rate capacity can be characterized to within one bit.
The channel conditions in \eqref{cond_m21_ic_opt_strat_Mk_1} and 
\eqref{cond_m21_ic_opt_strat_Mk_2} in Theorem \ref{thm_m21_ic_sum_cap_strat_Mk} 
for $K=3$ are summarized in Table \ref{table_many21_IC_results}. 
The sum-rate capacity in the low-interference regime, i.e., 
strategy $\mathcal{MI}1$ was proved in \cite{Veeravalli}.

\section{Conclusions}
 \label{sec_concl}
 
We considered the Gaussian many-to-one X channel with messages on all the
links. We formulated different transmission strategies and obtained sufficient 
channel conditions under which the strategies were either optimal or within a
gap from an outer bound. In the process, sum-rate capacity
was characterized in some sub-regions of the many-to-one 
X channel.  Subsequently, we identified a region in which the many-to-one 
X channel can be operated as a many-to-one interference channel without 
loss of sum-rate and further 
showed that in this region, {\bl the sum-rate capacity can be characterized 
to within a constant number of bits}. We next formulated transmission strategies for the 
Gaussian many-to-one interference channel and obtained channel conditions under 
which the strategies achieved sum-rate capacity. We also identified a region 
where sum-rate capacity can be characterized to within a constant number of bits. This region is larger than the region implied by the corresponding result for 
the Gaussian many-to-one X channel.

{\bl
We have restricted ourselves to the Gaussian many-to-one XC, since it is much harder to obtain exact sum-rate capacity results for the general fully connected $K \times K$ XC. 
The main difficulty lies in proving the decodability of intended message sets at the receivers for the various transmission strategies. For example, in case of the $K \times K$ many-to-one XC in standard form, we made use of Lemma \ref{lemma_dec_msg_rx_i} to show that 
under certain channel conditions,  $y_1$ is a degraded version of $y_i$ with respect to message $W_{1i}$ and hence $H(W_{1i}, W_{ii} \, | \, {\bf y}_i^n) \le 2 n \epsilon_n$.
We subsequently made use of this result in Theorem \ref{thm_low_interf} to prove the sum-rate optimality of strategy $\mathcal{M}1$, which involves using Gaussian codebooks and treating interference as noise. However, extending this result to the general $K \times K$ XC is not easy. It is not clear if identification of a smart genie is possible for this setting. In \cite{Geng_XC_ISIT, Geng_XC}, it has been shown that treating interference as noise (strategy $\mathcal{M}1$ in this paper) is optimal for the $K \times K$ XC for the sum-rate capacity up to a constant gap. It would be interesting to study the applicability of techniques used in \cite{Geng_XC_ISIT, Geng_XC} to analyze strategy $\mathcal{M}2$.
}

\end{document}

%% file: many_2_1_XC_sys_model.pdf_t
\begin{picture}(0,0)%
\includegraphics{many_2_1_XC_sys_model.pdf}%
\end{picture}%
\setlength{\unitlength}{3947sp}%
\begingroup\makeatletter\ifx\SetFigFont\undefined%
\gdef\SetFigFont#1#2#3#4#5{%
  \reset@font\fontsize{#1}{#2pt}%
  \fontfamily{#3}\fontseries{#4}\fontshape{#5}%
  \selectfont}%
\fi\endgroup%
\begin{picture}(3230,3920)(3018,-12937)
\put(3312,-12551){\makebox(0,0)[lb]{\smash{{\SetFigFont{9}{10.8}{\rmdefault}{\mddefault}{\updefault}{\color[rgb]{0,0,0}Tx $K$}%
}}}}
\put(5913,-12601){\makebox(0,0)[lb]{\smash{{\SetFigFont{9}{10.8}{\rmdefault}{\mddefault}{\updefault}{\color[rgb]{0,0,0}Rx $K$}%
}}}}
\put(3778,-12166){\makebox(0,0)[lb]{\smash{{\SetFigFont{8}{9.6}{\familydefault}{\mddefault}{\updefault}{\color[rgb]{0,0,0}1$K$}%
}}}}
\put(4906,-12709){\makebox(0,0)[lb]{\smash{{\SetFigFont{8}{9.6}{\familydefault}{\mddefault}{\updefault}{\color[rgb]{0,0,0}$KK$}%
}}}}
\put(6081,-12834){\makebox(0,0)[lb]{\smash{{\SetFigFont{8}{9.6}{\familydefault}{\mddefault}{\updefault}{\color[rgb]{0,0,0}$K$}%
}}}}
\put(3153,-12886){\makebox(0,0)[lb]{\smash{{\SetFigFont{8}{9.6}{\familydefault}{\mddefault}{\updefault}{\color[rgb]{0,0,0}$K$}%
}}}}
\put(5549,-12309){\makebox(0,0)[lb]{\smash{{\SetFigFont{8}{9.6}{\familydefault}{\mddefault}{\updefault}{\color[rgb]{0,0,0}$K$}%
}}}}
\end{picture}%

%% file: many_2_1_XC_Ktx_std_form.pdf_t
\begin{picture}(0,0)%
\includegraphics{many_2_1_XC_Ktx_std_form.pdf}%
\end{picture}%
\setlength{\unitlength}{3947sp}%
\begingroup\makeatletter\ifx\SetFigFont\undefined%
\gdef\SetFigFont#1#2#3#4#5{%
  \reset@font\fontsize{#1}{#2pt}%
  \fontfamily{#3}\fontseries{#4}\fontshape{#5}%
  \selectfont}%
\fi\endgroup%
\begin{picture}(5014,3012)(2222,-12029)
\put(2390,-11692){\makebox(0,0)[lb]{\smash{{\SetFigFont{8}{9.6}{\familydefault}{\mddefault}{\updefault}{\color[rgb]{0,0,0}1$K$}%
}}}}
\put(5549,-11334){\makebox(0,0)[lb]{\smash{{\SetFigFont{8}{9.6}{\familydefault}{\mddefault}{\updefault}{\color[rgb]{0,0,0}$K$}%
}}}}
\put(5913,-11626){\makebox(0,0)[lb]{\smash{{\SetFigFont{9}{10.8}{\rmdefault}{\mddefault}{\updefault}{\color[rgb]{0,0,0}Rx $K$}%
}}}}
\put(4021,-11086){\makebox(0,0)[lb]{\smash{{\SetFigFont{8}{9.6}{\familydefault}{\mddefault}{\updefault}{\color[rgb]{0,0,0}$K$}%
}}}}
\put(6054,-11868){\makebox(0,0)[lb]{\smash{{\SetFigFont{8}{9.6}{\familydefault}{\mddefault}{\updefault}{\color[rgb]{0,0,0}$K$}%
}}}}
\put(6687,-11889){\makebox(0,0)[lb]{\smash{{\SetFigFont{8}{9.6}{\familydefault}{\mddefault}{\updefault}{\color[rgb]{0,0,0}$KK$}%
}}}}
\put(7118,-9850){\makebox(0,0)[lb]{\smash{{\SetFigFont{8}{9.6}{\familydefault}{\mddefault}{\updefault}{\color[rgb]{0,0,0}1$K$}%
}}}}
\put(2356,-11978){\makebox(0,0)[lb]{\smash{{\SetFigFont{8}{9.6}{\familydefault}{\mddefault}{\updefault}{\color[rgb]{0,0,0}$KK$}%
}}}}
\put(3130,-11845){\makebox(0,0)[lb]{\smash{{\SetFigFont{8}{9.6}{\familydefault}{\mddefault}{\updefault}{\color[rgb]{0,0,0}$K$}%
}}}}
\put(3312,-11501){\makebox(0,0)[lb]{\smash{{\SetFigFont{9}{10.8}{\rmdefault}{\mddefault}{\updefault}{\color[rgb]{0,0,0}Tx$\,K$}%
}}}}
\end{picture}%